# Multimodal and Multiscale Interrogation of a Mechanically Tough Glass Forming Copper-Based Metal-Organic Framework

Mounir El Skafi[1], Guo-Qiang Li[2], Sophie R. Thomas[3], James D. Taylor[4], Mark Frogley[5], Gianfelice Cinque[5], Marc Pignitter[3], Michael Reithofer[3], Jia-Min Chin[3], Sebastian Henke[2], and Jin-Chong Tan[1*]

[1]Multifunctional Materials and Composites (MMC) Laboratory, Department of Engineering Science, Parks Road, Oxford, OX1 3PJ, University of Oxford, United Kingdom.

[2]Anorganische Chemie, Fakultät für Chemie und Chemische Biologie, Technische Universität Dortmund, Otto-Hahn Straße 6, 44227, Dortmund, Germany.

[3]Institute of Inorganic Chemistry, University of Vienna, Währinger Str. 42, 1090 Vienna, Austria.

[4]ISIS Neutron and Muon Source, Science and Technology Facility Council, UKRI, Rutherford Appleton Laboratory, Chilton, Didcot OX11 0QX, United Kingdom.

[5]Diamond Light Source, Harwell Campus, Chilton, Oxford OX11 0DE, United Kingdom.

[*]*Corresponding author*: jin-chong.tan@eng.ox.ac.uk

## Abstract

A copper-based metal-organic framework, $Cu(Im)_2$, was synthesized using a sol-gel process and subsequently melt-quenched into glass upon heating above 240 °C. In this paper, we present a multimodal, multiscale interrogation of the MOF nanocrystals and the resulting glasses. Structural characterization using X-ray diffraction, atomic force microscopy, and electron microscopy was performed to understand the morphology and size of the synthesized nanocrystals and glasses. Thermogravimetric analysis and differential scanning calorimetry were employed to understand the melting process of the crystals to form the glass. Synchrotron pair distribution function analysis was performed to verify that the framework structure is maintained upon melting, and nearfield infrared nanospectroscopy provided insight into the local chemical structure of the materials. The mechanical characterization *via* nanoindentation revealed that the resulting glasses exhibit an appreciably high elastic modulus (~10 GPa) and the highest fracture toughness ($K_{Ic}$ ~ 0.5 MPa $m^{1/2}$) yet reported for MOF glasses. This development is central to the emerging field of MOF-based materials processing and shaping, while upholding mechanical robustness and resistance to cracking.

## 1 Introduction

Metal-organic frameworks (MOFs) are hybrid organic-inorganic materials comprising metal clusters and organic linkers self-assembled in an extended three-dimensional (3D) framework. These materials are known for their intrinsic porosity and highly tunable physical and chemical properties [1, 2]. The remarkable potential of MOFs and their wide-ranging applications have attracted significant scientific interest, ultimately leading to the 2025 Nobel Prize in Chemistry being awarded to pioneering researchers in the field. Some of these applications include methane gas storage [3], photoluminescence [4], sensing [5], drug delivery [6, 7], triboelectric nanogenerators [8, 9], chemical warfare agent neutralization [10, 11], and water harvesting [12]. Many of these applications have already evolved into start-up companies, with many more expected to emerge in the future as the field continues to bridge the gap between academic research and industrial implementations [13].

MOFs are typically synthesized *via* solvothermal, electrochemical (thin film synthesis) [14, 15], sol-gel (MOF monoliths synthesis) [16, 17], and mechanochemical methods, the latter offering the advantage of solvent-free synthesis [18]. However, the challenge with these methods lies in either bulk shaping or the precise fabrication of robust MOF structures. Many attempts have been made to overcome this challenge and to better utilize MOFs for future applications, especially by incorporating MOFs into polymer matrices [19, 20] as in mixed-matrix membranes [21], porous foams [22], and 3D printed shapes [23, 24]. However, the challenge with such an approach is the propensity of pore blockage by polymer chains [25].

Another approach that has recently been explored to shape and fabricate pristine bulk MOF materials is to melt MOF powders and quench them into glasses. Melt-quenched MOF glasses are an emerging field that is still underexplored. MOF glasses are good candidates for promising applications, such as membranes for gas separation [26, 27], thin coatings [28], and micro-optics [29]. MOF glasses can be synthesized by different routes, including solvothermal methods and mechanochemical methods, as well as the incorporation of different metal nodes such as zinc, cobalt, copper [30], iron [31] and cadmium [32].

Since many MOFs undergo decomposition upon melting [33], only a limited number can be transformed into glasses [34]. Imidazolate (Im) and imidazolate-derived linkers are common linkers to form MOF glasses [35, 36]. However, the basic zinc-based zeolitic imidazolate framework (ZIF), $Zn(Im)_2$ commonly referred to as ZIF-4 melts at a temperature close to the decomposition temperature of the framework [35, 37], leading to the difficulty of having phase pure glasses without any chemical decomposition or impurities [38].

Bennett *et al*. [38], identified phase transformations of ZIFs and showed that the melting temperature can be lowered by using a mixed-linker ZIF-62 comprising imidazolate (Im) and benzimidazolate (bIm). Other studies have addressed ways to facilitate the melting of MOF glasses beyond using a mixture of metals and linkers, such as ionic liquids [39]. A recent study [40] demonstrates the promise of flux-mediated ligand exchange to facilitate melt-quenched hybrid glass formation of not only meltable ZIFs, but also of non-meltable carboxylate-based MOFs. In addition, the experiments done on MOF glasses have been complemented by computational studies using density functional theory (DFT) theory [41] and molecular dynamics (MD) simulations [42].

The mechanical properties of MOF glasses were explored mainly with ZIF-62 through indentation [43, 44], micropillar compression [45], and fracture under beam bending [46]. Nanoindentation reported that the elastic modulus of ZIF-62 is ~6 GPa and a hardness of ~650 MPa [44]. The fracture behavior has been explored for ZIF-62 glasses through Vickers microindentation [43] where the resulting cracks and the shear bands were observed without any quantitative understanding of the fracture toughness. To *et al*. [46] measured the fracture toughness of ZIF-62 glass by using a 3-point bending test. The reported fracture toughness (0.104 MPa $m^{1/2}$) is consistent with MD simulations; however, this fracture toughness value is relatively lower compared to dense hybrid polymorphs (~0.1−0.33 MPa $m^{1/2}$) [47] and other MOF monoliths studied to date [48, 49].

Nevertheless, the melting of ZIF-62 which initially has a lower porosity than other MOFs due to the presence of bIm [35, 37, 41] accompanied by the amorphization of the framework results in a reduction of porosity in the material [37, 50]. Various attempts have been proposed to introduce porosity to MOF glasses, such as the addition of porogens [51, 52] or by incorporating other MOF powders into the matrix (e.g., ZIF-7 [53] and ZIF-8 [54]). Tuffnel *et al.* discussed several possible routes to producing such composites, such as flux melting, blending, and crystal-glass composites [55].

Hitherto, among the attempts to synthesize meltable Cu-based MOFs and hybrid glasses, Xue *et al.* [30] reported a mechanochemical approach to synthesize a meltable copper imidazolate MOF with a $CuN_4$ coordination unit with a 3D network, and obtained its glass form although significant decomposition occurs since melting proceeded in parallel to decomposition at around 260 °C. Likewise, Watcharatpong *et al.* [56, 57] demonstrated that copper and an imidazolate variant can form a coordination polymer with a $CuN_2$ coordination unit with a 1D chain structure, which is meltable to yield ductile materials. Furthermore, a study in 2001 by Masciocchi *et al.* [58] reported several possible copper-imidazolate MOF structures and their corresponding topologies. In addition, Tian *et al.* [59]

have demonstrated a copper (I) imidazolate polymer to produce an anti-corrosive effect of imidazole with copper metal.

The present work introduces a previously unreported copper-imidazolate MOF structure, designated as $Cu(Im)_2$; $Im^-$ = imidazolate, whose X-ray diffraction pattern differs from all previously reported Cu-MOF frameworks (section 3.1). We provide a comprehensive structural characterization of both the crystalline MOF and its melt-quenched glass using a wide range of techniques. The $Cu(Im)_2$ glass was found to have a high stiffness, low melting temperature, and exhibits an exceptionally high fracture toughness compared with reported exemplars found in the literature [60].

## 2 Experimental Methods

### 2.1 Starting materials

Copper nitrate trihydrate ($Cu(NO_3)_2{\cdot}3H_2O$), imidazole (ImH), triethylamine (TEA), acetone ($CH_3COCH_3$), and methanol (MeOH) were purchased from Fisher Scientific and used as received.

### 2.2 Synthesis of $Cu(Im)_2$ and fabrication of glasses

0.482 g of copper nitrate trihydrate was dissolved in 9 mL of methanol in a glass vial. In another glass vial, 0.404 g of imidazole was dissolved in 9 mL of methanol yielding a metal to linker ratio of 1:3. 0.837 mL of triethylamine was added to the linker solution. The linker solution was added over the metal solution in a 50 mL centrifuge tube, and a purple gel was formed. The gel was sonicated for 5 minutes, and additional solvent was added up to 40 mL and the solution was centrifuged. The collected materials were washed three times in methanol and centrifuged, and the last wash was performed in acetone. The centrifuged monoliths were left to dry in air for 1 day. The collected chalky monoliths were melted at temperatures between 240 °C and 310 °C for 1-20 minutes. The melting process was performed in a preheated Ivoclar Vivadent furnace purged in a dry nitrogen atmosphere. The melted glasses were quenched by removing the materials from the furnace directly to ambient conditions. The molten glasses were not viscous when removed from the furnace and can be spread easily over the glass slide substrate. Images of the as-synthesized crystals in the form of a monolith and a sample of the melted glasses are shown in Figure 1a. The different samples of $Cu(Im)_2$ glasses will be referred to using the notation of $Cu(Im)_2$-temperature (°C)/time (min). For example, $Cu(Im)_2$-270/10 refers to the copper imidazolate glass prepared by melting the crystals in a preheated furnace at 270 °C for 10 minutes, followed by air quenching.

In this study, four $Cu(Im)_2$ glass samples, namely: 310/1, 270/10, 245/10, and 240/20, were prepared and observed to reach melting state under the specified heating conditions. The $Cu(Im)_2$-310/1 sample exhibited a green colour, while the other glass samples appeared brown. Among these, $Cu(Im)_2$-240/20 showed the darkest colour, likely due to prolonged heating causing decomposition. As a result, this sample was excluded from further detailed analysis. It was found from the structural characterization, especially the pair distribution function (PDF) analysis (section 3.3), that the samples of $Cu(Im)_2$-245/10 and $Cu(Im)_2$-270/10 have similar properties, and henceforth, the representative samples of $Cu(Im)_2$-245/10 were explored. Therefore, the $Cu(Im)_2$ crystals along with $Cu(Im)_2$-310/1 and $Cu(Im)_2$-245/10 were subsequently studied in depth.

## 2.3 Structural Characterization of the Crystals and Glasses

Powder X-ray diffraction (PXRD) was performed using a Rigaku Miniflex diffractometer employing the Bragg-Brentano geometry. It is equipped with a Cu K$\alpha$ radiation source, operating at 40 kV and 15 mA. The PXRD patterns of the as-synthesized $Cu(Im)_2$ crystals and melt-quenched glasses are presented in Figure 1b.

The attenuated total reflection Fourier transform infrared (ATR-FTIR) spectra in the mid-infrared (MIR) region were collected using a Thermo Scientific Nicolet iS10 spectrometer equipped with a diamond ATR module. Synchrotron radiation far-infrared (SR-FIR) ATR and (micro)spectroscopic measurements were conducted at beamline B22 in the Diamond Light Source (DLS) (Supporting Information (SI), section 2).

Micro-Raman spectroscopy was performed using a Renishaw inVia Raman microscope equipped with a 785 nm laser and 1200 lines per mm grating. The data was collected under a 50× objective lens, using 1% laser power (of maximum 300 mW), a sample exposure time of 15-20 seconds, and 20-50 accumulations per spectrum.

$^1$H NMR spectra were recorded on a Bruker DPX-400 spectrometer using digested samples. A mixture of hexadeuterodimethyl sulfoxide, DMSO-$d_6$ (600 μL), and deuterium chloride in deuterium oxide, $DCl/D_2O$ (35 wt%, approx. 25 μL), was used as solvents. The spectra were processed with MestReNova software. Chemical shifts (ppm) were referenced to the residual proton signal of DMSO-$d_6$ at 2.50 ppm. The signal of the residual protons of $DCl/D_2O$ appears at variable chemical shifts depending on its concentration.

Simultaneous thermogravimetric analysis and differential scanning calorimetry (TGA-DSC) measurements were conducted on a SDT650 instrument (TA Instruments) under a constant $N_2$ flow of 100 mL min$^{-1}$ at a heating rate of 10 °C min$^{-1}$. Samples were measured in alumina ceramic crucibles.

DSC measurements were carried out on a DSC25 instrument (TA Instruments) under $N_2$ flow of 50 mL $min^{-1}$ with a heating/cooling rate of 10 °C $min^{-1}$. Hermetically sealed aluminium crucibles (type: 901683.901/901684.901) with a hole pinched in the lid were used.

Thermogravimetric analysis – mass spectrometry (TGA–MS) measurements were performed using a Setaram Setsys Evolution TG–DSC instrument coupled to a Pfeiffer Vacuum OmniStar GSD 350 mass spectrometer. Samples were heated in 90 μL alumina crucibles under an argon flow (99.999 vol%; 16 mL $min^{-1}$) at a heating rate of 10 °C $min^{-1}$. A 10 min isothermal hold at 140 °C was included to remove residual solvent before further heating.

Volumetric and surface area characterization were performed on a Quantachrome Autosorb iQ-Chemi instrument. The nitrogen adsorption and desorption isotherms at 77 K were measured on the crystals and the crushed glasses. More details are provided in the Supplementary Information (SI, section 3).

Scanning electron microscope (SEM) imaging of the nanocrystals was performed using the Tescan Lyra3 under 1 kV and a short working distance of 4 mm in secondary electron (SE) imaging mode. Retractable scanning transmission electron microscopy (R-STEM) was performed employing the same Tescan system, by drop casting the crystals using acetone onto the TEM grid and operating the system at 30 kV and a working distance of 5 mm. The imaging of the residual indents and morphology of fracture surfaces was performed on uncoated samples, employing a tabletop Hitachi TM3030Plus SEM under low vacuum.

High-resolution transmission electron microscopy (HR-TEM) measurements were carried out in the Electron Microscopy Facility at the Institute of Science and Technology Austria (ISTA), using a S/TEM JEOL JEM-2800 instrument with an accelerating voltage of 200 keV, equipped with a CMOS TEM camera, TemCam-XF416. All samples were prepared by grinding the glass to a fine powder, suspending them in acetone, and drop-casting onto 200-mesh copper grids coated with a carbon film. The grids were dried in a 70 °C oven for at least one hour before imaging. The HR-TEM images were processed using the EM Measure software.

Continuous-wave electron paramagnetic resonance (CW-EPR) spectra were recorded at room temperature using an X-band Bruker Elexsys-II E500 EPR spectrometer (Bruker BioSpin GmbH, Rheinstetten, Germany) equipped with a high-sensitivity SHQE1119 cavity. Measurements were performed at a microwave frequency of 9.86 GHz with a modulation frequency of 100 kHz. The center field was set to 3500 G with a sweep width of 5000 G and a sweep time of 38.1 s. The modulation amplitude and microwave power were set to 7 G and 0.63 mW, respectively. Spectra were acquired using 1024 points and averaged over 10 scans. The spectra were processed and analyzed using Bruker Xepr software.

The surface height topography of the crystals was measured by atomic force microscopy (AFM) as implemented in a scattering-type scanning near-field optical microscopy (s-SNOM) instrument (Neaspec GmbH) under the tapping mode. An Arrow-NCR probe was used, with a nominal tip radius of < 10 nm, a nominal stiffness of 42 N/m and resonant frequency of 285 kHz.

### 2.4 Nanoindentation measurements

Instrumented nanoindentation was performed using a KLA Instruments iMicro system, employing the continuous stiffness measurement (CSM) method. A Berkovich (3-faced pyramidal) diamond indenter tip was used for elastic modulus ($M$) and hardness ($H$) measurements on the glass samples that were epoxy mounted (Struers Epofix) and carefully polished to a surface roughness of below 1 μm. An average of 25 indentations were carried out with the CSM method to determine the values of indentation modulus ($M$ obtained by letting the Poisson's ratio of sample, $\nu_s = 0$ [60]) and hardness ($H$ = maximum load divided by projected contact area under load) as a function of the surface penetration depth for different glass samples. The typical maximum indentation depth was set to 2000 nm. Thermal drift measurements were performed at 90% unload, held for a period of 60 s. The Oliver and Pharr (O&P) method was applied to extract the values of $M$ and $H$ from the load-displacement ($P-h$) curves; see assumptions used and further details in the SI (section 4).

### 2.5 Pseudoheterodyne (PsHet) nearfield infrared imaging and point spectroscopy (PSP)

Nearfield PsHet nanoimaging with the Piano wOPO (widely tunable optical parametric oscillator) infrared laser (Stuttgart Instruments) was performed on the Neaspec s-SNOM under tapping mode. An electrically conductive Arrow-NCPt platinum iridium-coated tip was used, with a nominal tip radius of below 25 nm, resonant frequency of 285 kHz, and a tapping amplitude of ~65 nm during scanning. Background calibration was conducted using the TGQ-1 silicon/$SiO_2$ standard on the silicon region. The glass samples were imaged at 1098 cm$^{-1}$ corresponding to the vibrational mode of the imidazolate linker in the $Cu(Im)_2$ framework, and at 1050 cm$^{-1}$ as a control scan having no vibrational modes of $Cu(Im)_2$.

s-SNOM point spectroscopy (PSP) measurement [61] was performed over a frequency range of 700−1600 cm$^{-1}$. Each PSP spectrum was collected from a single scan with an integration time of 60 ms per wavenumber, where the laser source provides individual wavenumbers sequentially, the near-field response on the material was detected accordingly. The spectral and spatial resolutions of PSP are 1 cm$^{-1}$ and ~20 nm, respectively. The collected spectra include different demodulation harmonics of the optical amplitude (A) and phase (P) signals, each representing a different nearfield interaction volume in scanning depth. The spot size for the second harmonic (O2) is around 25 nm, O3 is around 20 nm, and O4 is around 15 nm. Normalization of the different optical harmonic signals (e.g., O3P − O2P) allows the elimination of unwanted far-field effects and geometrical artefacts, thus giving an improved nearfield characterization of the nanocrystals and glasses [62].

# 3 Results and Discussion

## 3.1 Structural and textural analyses

The $Cu(Im)_2$ crystals which were synthesised through a sol-gel process and the $Cu(Im)_2$ glasses which were subsequently melted in the furnace under nitrogen gas purging are shown in Figure 1a. The as-synthesized crystals are obtained in the form of a chalky monolith of pale purple color, and the representative samples of the melt-quenched glasses which were prepared on a glass slide exhibit a green color for $Cu(Im)_2$-310/1 and a brown colour for $Cu(Im)_2$-245/10. PXRD patterns (Figure 1b) confirm that the resulting $Cu(Im)_2$ is crystalline, with two sharp low-angle Bragg peaks located below an angle of $2\theta = 7°$. As expected, the glass samples of ($Cu(Im)_2$-310/1 and $Cu(Im)_2$-245/10) do not show any peaks in the PXRD, indicating that the resulting glasses are amorphous. The absence of Bragg peaks between 30° and 40° in both glass samples indicates that the melting process does not result in the formation of copper oxides, and the observed change of sample colour into brown is due to the partial decomposition of the framework when preparing $Cu(Im)_2$-245/10, due to a longer heating time. In contrast, $Cu(Im)_2$-310/1 glass is green possibly due to the change in the coordination of the framework.

The PXRD pattern of $Cu(Im)_2$ crystals does not match any of the reported copper(II) imidazolate frameworks in the Cambridge Structural Database (CSD), suggesting that the synthesized material is a new framework. SEM imaging of the crystals (Figure 1c) reveals that the crystals of $Cu(Im)_2$ are rod like. AFM imaging of the crystals shown in Figure 1d–g reveals defective, twinned nanocrystals with a fine needle-like morphology, comprising nanostructures that are consolidated into bundles (Figure 1e,g). The as-synthesized crystals have lengths of around 2 μm and a diameter below 100 nm. To further confirm the crystal morphology, retractable scanning transmission electron microscopy (R-STEM) imaging was performed on the nanocrystals; see images shown in SI Figures S1-S3.

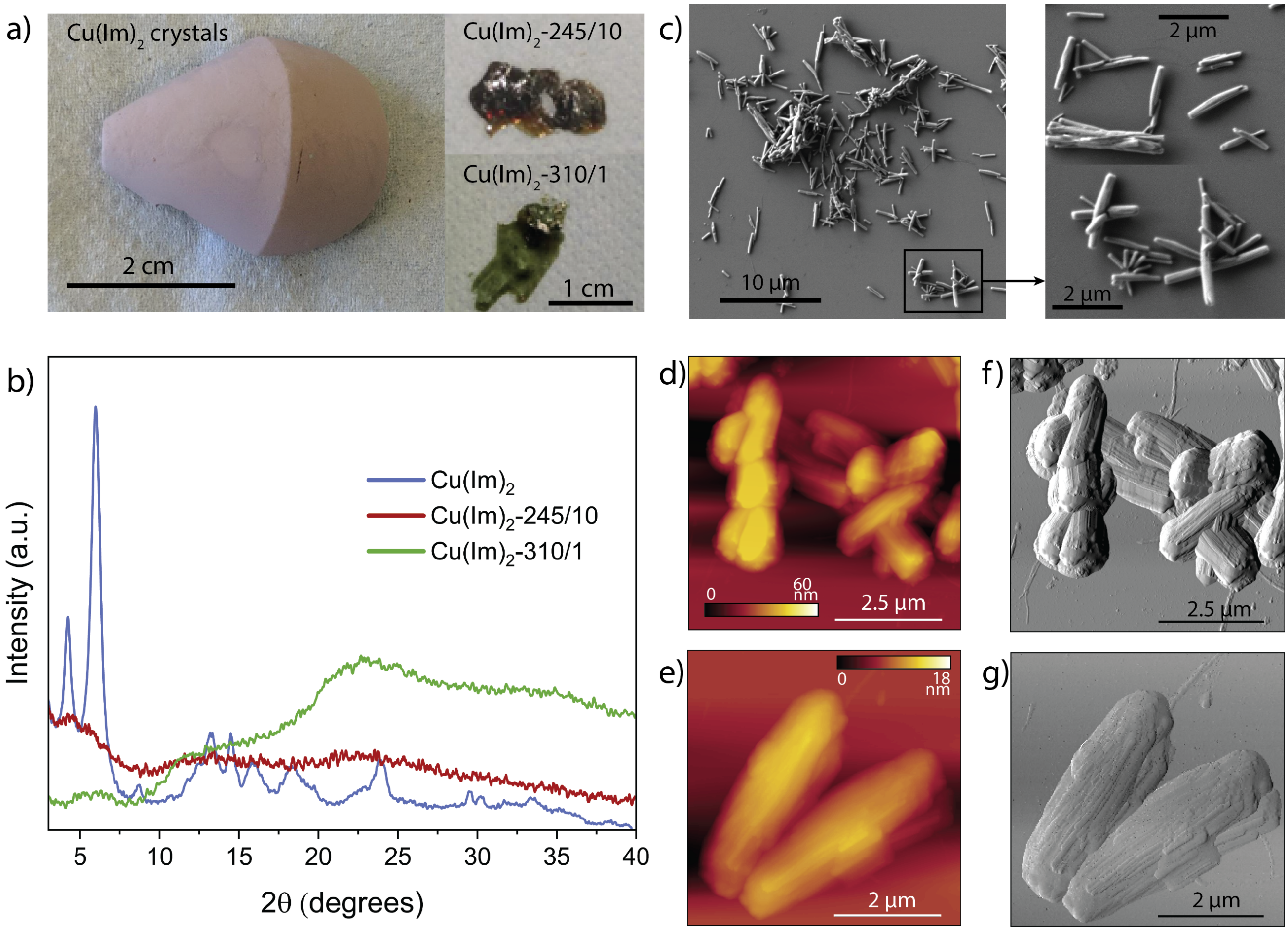

Figure 1. (a) Representative monolith and melt-quenched glass samples of $Cu(Im)_2$ employed in this study, showing the chalky monolith in pale purple, $Cu(Im)_2$-310/1 in green (bottom right), and $Cu(Im)_2$-245/10 in brown (top right). Note: the labels designate the processed sample as $Cu(Im)_2$-temperature (°C)/time (min). (b) PXRD patterns of $Cu(Im)_2$ and the resulting glasses prepared under different temperature and heating time. (c) SEM images for the single crystals of $Cu(Im)_2$ drop casted on a silicon substrate. (d,e) AFM height topography images of the as-synthesized nanocrystals of $Cu(Im)_2$ (drop-casted on a flat silicon substrate). (f,g) AFM mechanical amplitude images of the as-synthesized nanocrystals of $Cu(Im)_2$ (drop casted on a flat silicon substrate).

HR-TEM imaging on $Cu(Im)_2$ crystals and glasses was performed, and the results are presented in Figure 2. The imaging aimed to reveal the nanoscale transformations of the different $Cu(Im)_2$ glass samples from a crystalline lattice to an amorphous glass. Figure 2a shows the $Cu(Im)_2$ single crystals that were drop casted on the TEM grid, and Figure 2b reveals the ordered lattice structure of these crystals, where the parallel oblique lines suggest an ordered structure that agrees with the PXRD analysis. The *d*-spacing of the lattice was found to be 2.98 Å by applying inverse fast Fourier transform (FFT). Looking at the glasses, $Cu(Im)_2$-310/1 shows that amorphization occurs, however, small nanodomains with some order are observed at *ca*. 1-2 nm (Figure 2c), despite the PXRD showing an amorphous structure for the same sample (Figure 1b). These ordered nanodomains are too small and localized to be detected by a benchtop PXRD and thus, the overall structure is amorphous, and their presence is likely due to the very rapid glass-forming process. This is clearly depicted for the other glass samples $Cu(Im)_2$-245/10 and $Cu(Im)_2$-270/10 in Figure 2d-e, respectively, which show amorphized glasses and no signs of decomposition suggesting that the ordered nanodomains in $Cu(Im)_2$-310/1 are due to the short melting time. Finally, Figure 2f shows the sample of $Cu(Im)_2$-240/20 and reveals clearly that decomposition is prevalent. Although the preparation of $Cu(Im)_2$-240/20 glass in bulk yields a freestanding material, the resultant product is in fact a decomposed form of the MOF, as determined *via* PDF analysis (section 3.3).

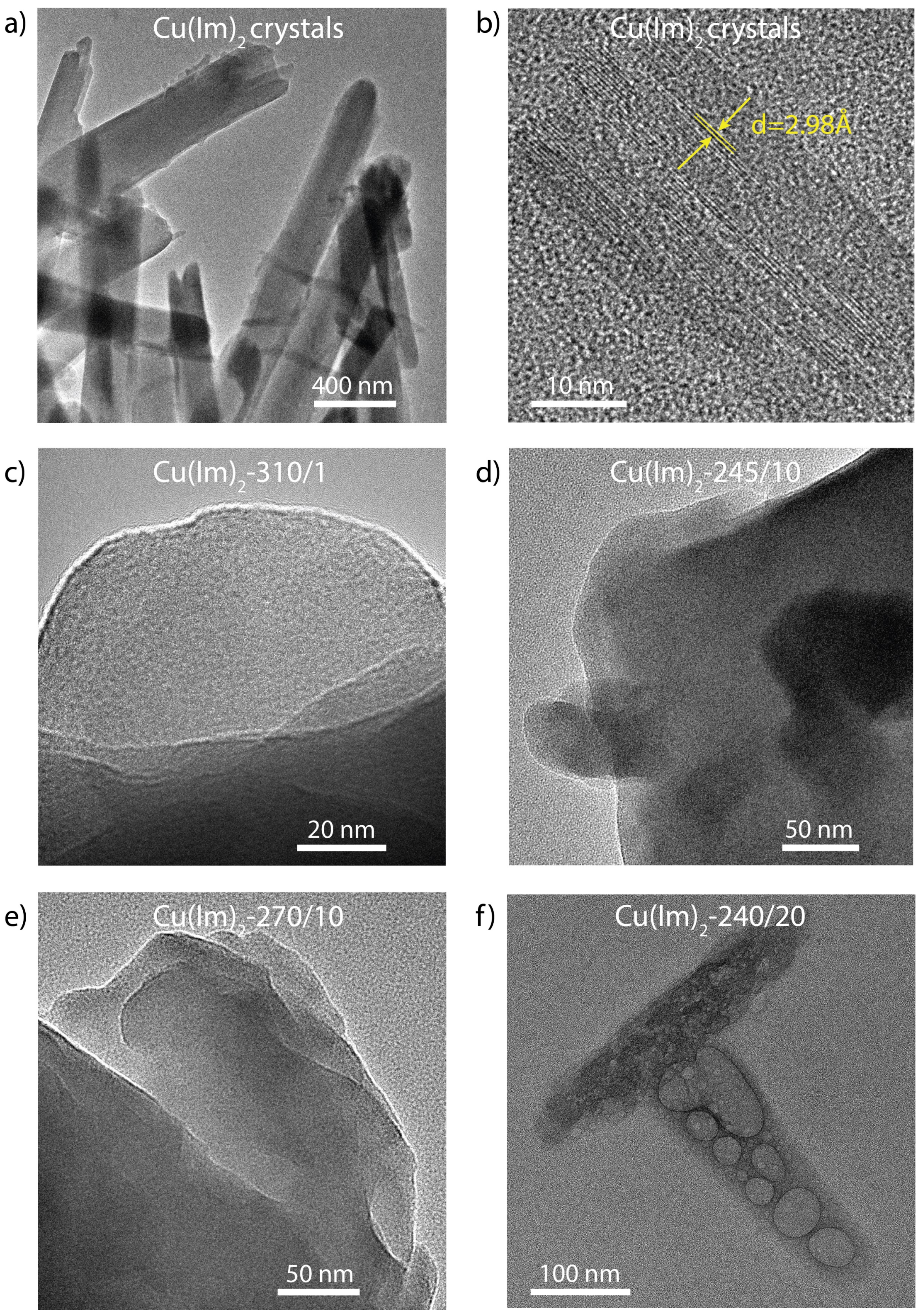

Figure 2. HR-TEM images of (a) $Cu(Im)_2$ crystals, (b) a magnified view of the crystals showing the ordered lattice planes with a *d*-spacing of 2.98 Å, (c) $Cu(Im)_2$-310/1 glass showing nanodomains of an ordered lattice, (d) $Cu(Im)_2$-245/10 and (e) $Cu(Im)_2$-270/10 glasses, and (f) $Cu(Im)_2$-240/20 showing the decomposed material.

### 3.2 Understanding the Melting of $Cu(Im)_2$ Monoliths through Thermal Analysis

In order to understand the melting mechanism of the $Cu(Im)_2$ crystals, thermogravimetric analysis (TGA) and differential scanning calorimetry (DSC) were performed. Figure 3a shows the coupled TGA-DSC scans which identifies partial decomposition of the framework at around 240 °C accompanied by an exothermic DSC peak, and all TGA weight losses below 150 °C are attributed to trapped solvent in the crystals. DSC was performed separately with two upscans as shown in Figure 3b, reaching 250 °C, aiming to identify an endothermic signal associated with melting before the exothermic thermal decomposition event. It can be seen in Figure 3b that melting occurs at ~225 °C and the corresponding endothermic signal overlaps with a subsequent exothermic signal assigned to decomposition. The second upscan shows a slight convex curve at around 200 °C, which is close to the melting temperature of the crystals. An earlier glass transition temperature for the $Cu(Im)_2$-310/1 glasses can be seen at around 110 °C while varying the DSC heating rate (Figure S4).

The TGA-DSC results confirmed that melting $Cu(Im)_2$ crystals is possible. We note that the preparation of the glass samples in furnace happens at a higher heating rate than the TGA-DSC experiment, which is inevitable because the crystals are directly inserted into the preheated furnace chamber and kept for a very short time in comparison to the controlled TGA-DSC experiments. These preparation parameters allow the melting and formation of glass while limiting the decomposition of the framework structure.

### 3.3 Probing the Structure of $Cu(Im)_2$ through Synchrotron Pair Distribution Function (PDF)

In order to verify that the resultant material is a framework material which has amorphized into glass and not a completely decomposed compound, synchrotron pair distribution function (PDF) analysis was performed on all the prepared samples. Figure 3c shows the local structure of the $Cu(Im)_2$ framework and the extended PDFs for the crystals and glasses of $Cu(Im)_2$-245/10, $Cu(Im)_2$-270/10, $Cu(Im)_2$-310/1, and $Cu(Im)_2$-240/20. Figure 3d shows a magnified view of the PDFs of the crystalline and glassy phases, revealing that the crystalline $Cu(Im)_2$ shows a clear signature of a ZIF structure composed of imidazolate-bridged Cu ions. The $Cu(Im)_2$ crystals show a distinct copper nitrogen coordination at 1.97 Å for the Cu−N bond. For the glasses, the copper nitrogen coordination at 1.97 Å is shifted to 1.87 Å suggesting that the coordination is still present, but distortion to the original structure occurs. The peak at 2.2 Å in the glasses represents the diagonal non-bonded atoms within the imidazole ring, namely C···N, C···C, and N···N. Furthermore, the results of NMR (SI Figures S5 and S6) on the crystals and $Cu(Im)_2$-240/20 show that a copper imidazolate structure is preserved upon melting.

In order to understand the structural transformation of the $Cu(Im)_2$ crystals into glass, TGA analysis coupled with mass spectroscopy (TGA-MS) (SI Figure S7) and electron paramagnetic resonance (EPR) (SI Figure S8) were performed. From TGA-MS, around the melting temperature where the first decomposition stage occurs, the crystals are losing nitrogen products, namely NO along with $CO_2$, CO and water. Therefore, we reasoned the melting can be associated with decomposition where a redox reaction would reorganize the $Cu(Im)_2$ framework into a Cu(Im) system, through the elimination of an Im radical which is released as the species determined by mass spectrometry (SI Figure S7). This agrees with the mass loss of ~30% above the melting temperature in the TGA, which aligns well with the theoretical mass loss (~34%) associated with the reduction of $Cu(Im)_2$ to Cu(Im), and the loss of one imidazole ligand. Also, based on the shift in the bond length from PDF at 1.97 Å, the bond length of the glasses agrees with a copper (I) imidazolate-based polymer [59]. The imidazolate radical will be unstable and decompose into gaseous product along with other radicals to form carbonaceous species as detected in TGA-MS (SI Figure S7). These carbonaceous species cannot be detected by NMR, but likely are the explanation for the colour change observed in the melt-quenched glass samples (Figure 1a). However, the faster the melting process is, the less decomposition occurs, and this explains why the structure of $Cu(Im)_2$-310/1 is more preserved.

To verify if Cu(II) is present in the glass samples, EPR was performed (SI Figure S8), and the observed resonating peaks in the crystals at $g = 2.16$ and $g = 1.94$ indicate an axially symmetric Cu(II) centre, consistent with an axially compressed pseudo-octahedral geometry. Upon heating, the marked attenuation of the principal EPR signal from Cu(II) indicates a substantial change in the coordination and magnetic environments of the copper centres. Amongst the different glasses, the least attenuation was observed in $Cu(Im)_2$-310/1 although it is also largely attenuated. This may result from stronger antiferromagnetic exchange between neighbouring Cu(II) ions as the network rearranges, causing some Cu(II) populations to become EPR-silent or their resonances to broaden considerably. The increased structural heterogeneity of the amorphous glass material may further broaden the signal through a distribution of Cu geometries and $g$-values. However, the partial reduction of Cu(II) to the EPR-silent Cu(I) is another supporting evidence which agrees with the decomposition step observed in TGA. The factor that determines exactly how much decomposition happens is the speed of the process, where a faster melting process prevents further decomposition yet allows reorganization of the initial framework to form a glass.

Taken together, the EPR and PDF results show that heating substantially reorganizes the principal Cu coordination and magnetic environment during amorphization. The observations are consistent with the distorted and increasingly heterogeneous copper coordination, accompanied by an altered Cu(II)–Cu(II) magnetic coupling. Partial formation of EPR-silent Cu(I), however, cannot be excluded.

From this analysis, we confirm that the melt-quenched materials are MOF glasses with the most promising samples being $Cu(Im)_2$-310/1, followed by $Cu(Im)_2$-270/10 and $Cu(Im)_2$-245/10. For the remaining of the analysis, we limit our characterisation to three samples: the pristine $Cu(Im)_2$ crystals as a reference to all processing performed; $Cu(Im)_2$-310/1 as the most promising sample showing the least decomposition, and with the closest coordination similarity to the pristine sample; and $Cu(Im)_2$-245/10 being a glass sample formed at a different processing temperature which is similar to $Cu(Im)_2$-270/10.

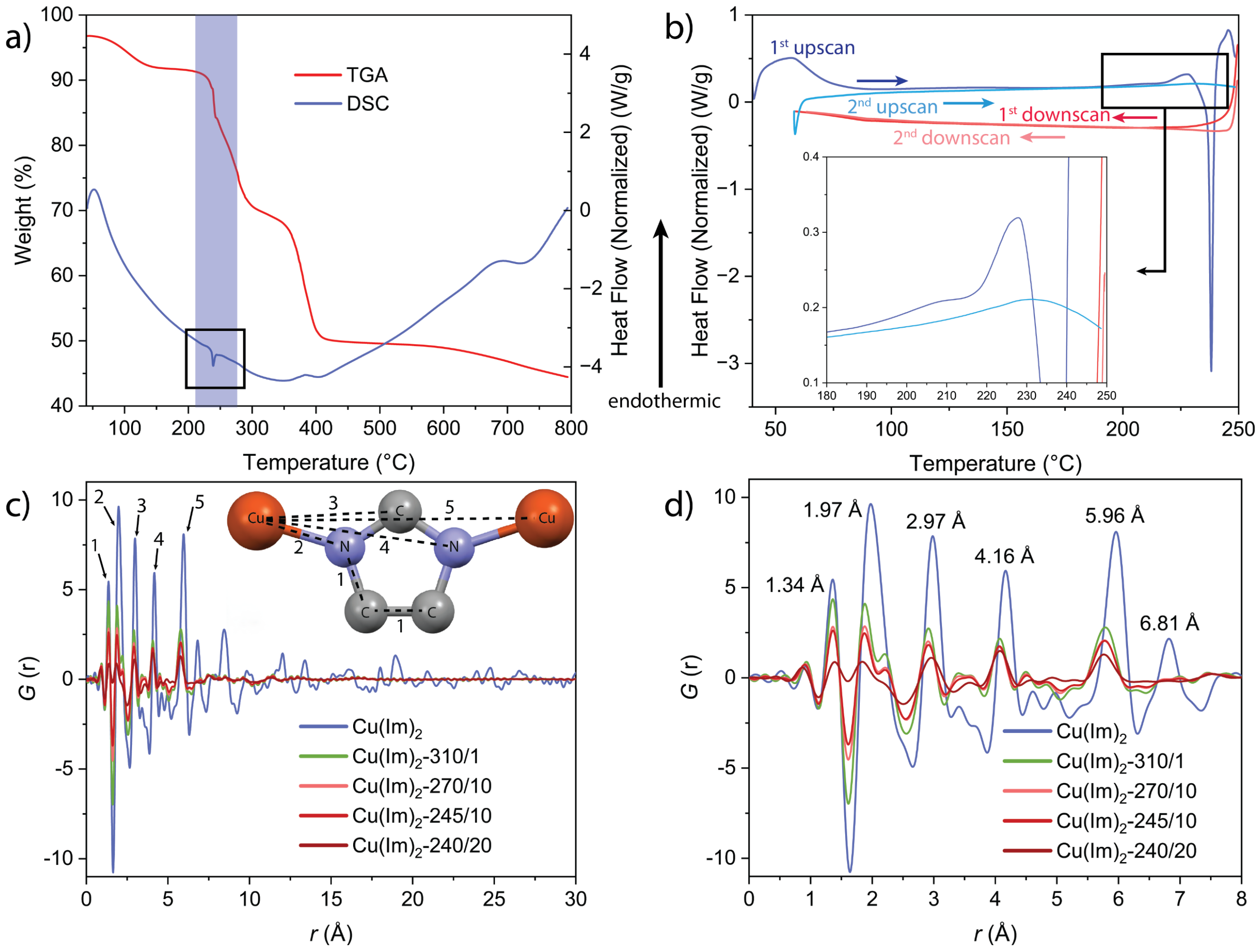

Figure 3. (a) Coupled TGA-DSC measurements performed on the $Cu(Im)_2$ crystals. (b) DSC of $Cu(Im)_2$ crystals with two upscans and downscans. (c) Representation of the local structure of $Cu(Im)_2$ and synchrotron PDFs of crystalline and glassy $Cu(Im)_2$ prepared using different melting parameters. (d) Magnified view of the synchrotron PDFs showing notable structural changes upon heat treatment.

### 3.4 Spectroscopic analysis of $Cu(Im)_2$ crystals and glasses

Figure 4a presents the ATR-FTIR spectra for $Cu(Im)_2$ crystals and glasses normalized to the 1085 $cm^{-1}$ band. For $Cu(Im)_2$ crystals, the ATR-FTIR shows sharp vibrational modes with the most prominent band at 1085 $cm^{-1}$ assigned to the C−N stretching of the imidazolate (Im) ring. For the two glass samples, $Cu(Im)_2$-310/1 and $Cu(Im)_2$-245/10, the ATR-FTIR spectra are almost identical. The difference in spectra between the crystals and glass is the attenuation and broadening of all vibrational modes, and this is attributed to the amorphization and redistribution of the framework structure. Nevertheless, the position of the bands in all the samples is retained in the sample. An additional vibrational mode at 1605 $cm^{-1}$ appears for both glass samples, and this band refers to the conjugated skeletal stretch (ring C=C/C=N-like) in the imidazolate ring. This absorption band appears when melt quenching creates stronger conjugated/altered linker environments (defects, altered coordination), and this band is more prominent in $Cu(Im)_2$-245/10 due to higher exposure time under heat. This suggests altered coordination due to the amorphization of the framework by melting. The vibrational mode at around 950 $cm^{-1}$ which refers to the C−N−C stretching of the Im ring is highly attenuated in the glass samples, and it is significantly broadened in $Cu(Im)_2$-245/10 due to prolonged heating. This happens because the formation of glass shifts this region into overlapping sub-bands, leading to partial loss of intact Im linker environments and thus reducing intensity. The extended ATR-FTIR spectra for the same samples are shown in Figure S9, revealing traces of imidazole identified from the low-intensity band at 3130 $cm^{-1}$. The presence of imidazole is likely to be residual from washing the crystals and remained when preparing the glass.

In order to understand the effect of pressure on the interaction between copper and nitrogen in the framework, high-resolution synchrotron ATR far-infrared (FIR) measurements to probe the Terahertz (THz) vibrational modes were performed at B22 in Diamond Light Source (DLS). Samples comprised 13-mm pellets prepared from $Cu(Im)_2$ crystals compressed at 1 ton and 5 tons of load for 5 minutes in a manual hydraulic press, and a melt-quenched $Cu(Im)_2$-310/1. Figure 4b shows the FIR spectra of the different samples, revealing the Cu-N stretching detected at ~10 THz (345 $cm^{-1}$). The pelleting pressure was found to broaden the peak and shift it to a lower wavenumber, which indicates the framework is partially amorphized and/or the structure is strained. However, upon melting and melt quenching of the crystals, we found that the Cu-N coordination mode at ~10 THz is preserved and exhibits an enhanced intensity compared to the original crystals. We reasoned that the melting process reorganizes the copper atoms in the framework to have a higher oscillation between the copper and nitrogen increasing the intensity of the vibrational band. Furthermore, synchrotron FIR spectroscopy was performed under an IR microscope on the $Cu(Im)_2$ crystals, the micro-FIR spectrum shown in Figure S10 agrees with the ATR-FIR of the crystals. Further details about synchrotron spectroscopy at DLS B22 can be found in SI section 2.

In addition, micro-Raman spectroscopy was performed on the same set of samples, and the collected spectra are presented in Figure 4c. The Raman spectra for the $Cu(Im)_2$ crystals exhibit sharp and well-defined Raman-active modes, with the most prominent band detected at around 975 $cm^{-1}$ attributed to the breathing and symmetric stretching modes of imidazolate rings. This band along with all bands in the glass samples are strongly attenuated primarily due to the amorphization of the framework. It can be seen that the effects on amorphization are more pronounced in Raman compared to ATR-FTIR. The extended Raman spectrum for $Cu(Im)_2$ crystals is shown in Figure S11 and does not reveal any additional bands beyond 1500 $cm^{-1}$.

### 3.5 Volumetric and surface area assessment by nitrogen sorption

Nitrogen sorption measurements were performed at 77 K to investigate the porosity of $Cu(Im)_2$ crystals and to assess the effect of melting and glass formation on the porosity of the framework. Figure 4d shows the Type-I/II like isotherms established from the volumetric adsorption of nitrogen per unit mass. The surface areas determined by the Brunauer–Emmett–Teller (BET) method are presented in Table S1. Sorption data revealed that $Cu(Im)_2$ crystals retain a low surface area of ~112 $m^2/g$, suggesting that $Cu(Im)_2$ crystals are relatively dense compared to more porous MOFs, and have a similar surface area higher than ZIF-zni but lower than ZIF-4 [63]. The $Cu(Im)_2$ glasses exhibit a considerably lower surface area of ~20-30 $m^2/g$, which is expected as it is seen in other MOF glasses such as ZIF-62 [37]. However, there is possibility for increasing the porosity of dense systems by adding other non-meltable MOFs into the glass [51], in order to obtain robust structures that maintains porosity from loose MOF crystals.

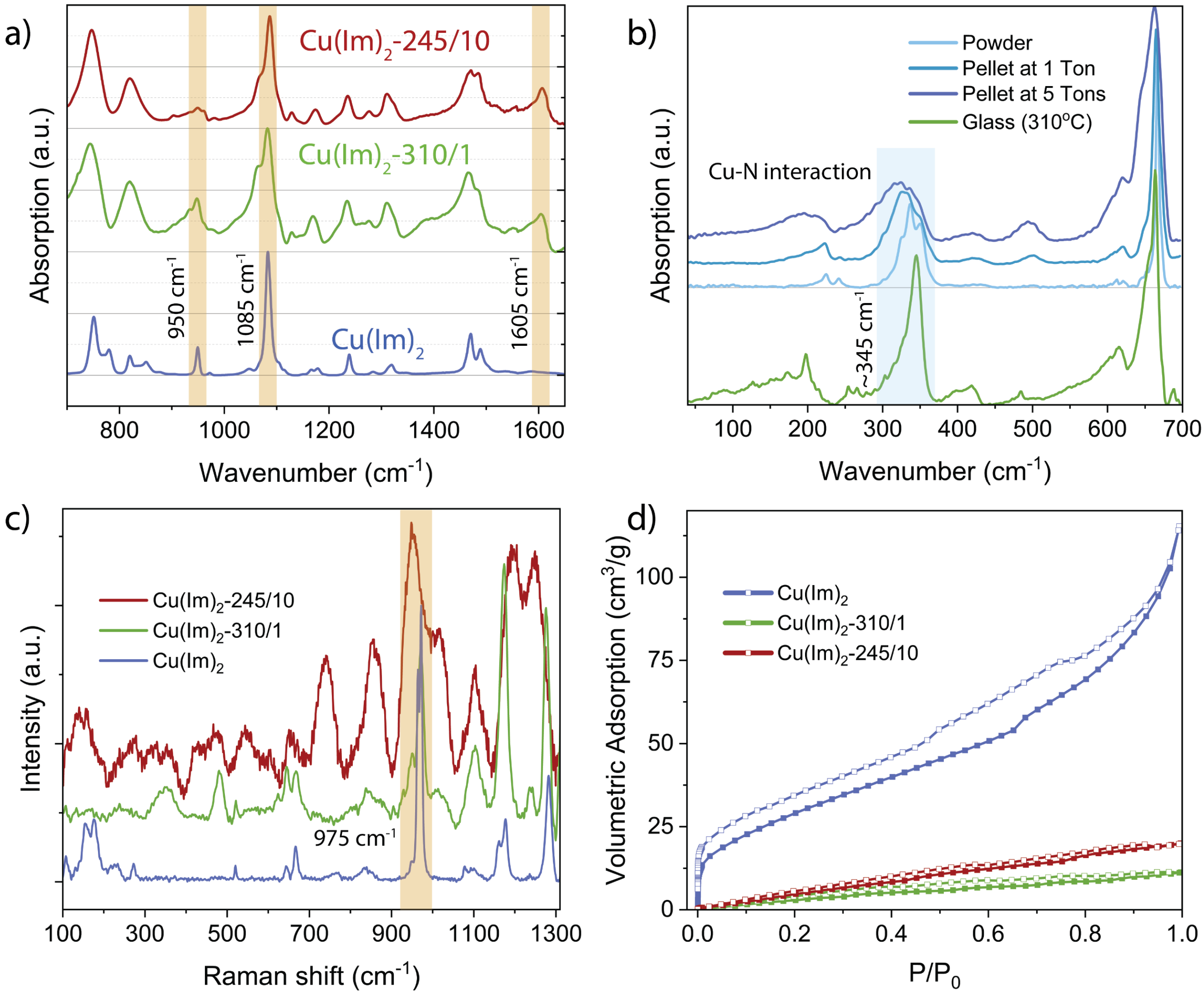

Figure 4. (a) ATR-FTIR spectra for the $Cu(Im)_2$ crystals and glasses. (b) Synchrotron ATR-FIR spectra for $Cu(Im)_2$ crystals pelleted at different loads and $Cu(Im)_2$-310/1 glass. (c) Micro-Raman spectra for $Cu(Im)_2$ crystals and glasses. (d) Nitrogen adsorption and desorption isotherms of $Cu(Im)_2$ crystals and glasses at 77 K. Closed and open symbols representing adsorption and desorption test segments, respectively.

### 3.6 Nearfield pseudoheterodyne (PsHet) infrared imaging and point nanospectroscopy (PSP) of the crystals

The nanoscale understanding of the local structure of $Cu(Im)_2$ crystals and glasses was acquired by nearfield infrared characterization *via* a scattering-type scanning nearfield optical microscope (s-SNOM) equipped with pseudoheterodyne (PsHet) nanoimaging and point nanospectroscopy. Two adjacent crystals probed using the AFM topographic mode are shown in Figure 5a. The infrared wOPO laser was tuned to 1098 cm$^{-1}$, corresponding to the fingerprint wavenumber of the imidazolate linker (Figure 4a). Figure 5b highlights the IR absorption region over the scanned area in red, hence highlighting the crystals with a strong contrast against the (reflective) silicon background. In order to verify that the contrast is due to the presence of the Im linker on the crystals and not a geometric artefact, the same region was probed at 1050 $cm^{-1}$ (a control wavenumber that does not correspond to any vibrational modes in the crystals), and the results verified that there is no absorption over the crystals (Figure 5c). To further understand the accuracy of the nearfield signal over the $Cu(Im)_2$ crystals, point nanospectroscopy (PSP) was performed on the crystals and the local absorption spectra were collected under a spatial resolution of ~20 nm [64]. Different optical harmonics were observed in Figure 5d, where the higher demodulation harmonics signify a higher nearfield signal (i.e. O4 > O3 > O2), and the difference between the consecutive harmonics allows us to remove any unwanted far-field signals due to scattering or reflection from the sample surroundings [62]. Although no background bands are expected to appear in the spectra, and the different harmonics show high similarity by capturing all vibrational modes, the difference in optical phase signals between the different harmonics enabled elimination of sample background effects and yielded an improved spectrum with a flatter baseline. The vibrational bands captured through point spectroscopy are in agreement with the spectra by ATR-FTIR in Figure 4a, verifying that the bulk structure is identical to the local structure of individual crystals. There is a shift in all bands by ~10 $cm^{-1}$ between the ATR-FTIR spectra and point spectroscopy attributed to nearfield effects [61], but this is consistent with all nearfield spectra.

Additional PsHet nanoimaging with different harmonics of the optical phase signal is shown in Figures S12-S14. It was found that that the fourth optical harmonics O4, the most nearfield signal, has less sensitivity to geometric changes and captures the local chemical variation over the scanned area (rather than geometric artefacts). Accordingly, it is evident from Figure S12 that the O2 image shows a smooth transition between the edges of the crystal and substrate compared to the image of the O4 signal, which has more contrast at the edges of the crystal because of the higher nearfield confinement of the signal compared to O2 that captures more signal from the surroundings. This is further improved by taking the difference in harmonics, for example O3P − O2P and O4P − O3P (Figures S12-S14) to reduce far-field effects, but this approach may compromise the signal intensity. Therefore, in the rest of our analysis, we will focus on the fourth harmonics of optical phase O4P.

To further understand the nanoscale structure of the glasses, point nanospectroscopy was performed on $Cu(Im)_2$-310/1 and $Cu(Im)_2$-245/10 and compared with $Cu(Im)_2$ crystals. The fourth harmonic optical phase signals (O4P) are presented in Figure 5e. The PSP results show good agreement with the spectra from ATR-FTIR (Figure 4a) as well in terms of peak broadening, especially in the range between 750 $cm^{-1}$ and 900 $cm^{-1}$. Also, the band at 960 $cm^{-1}$ is highly attenuated in $Cu(Im)_2$-245/10 due to prolonged heating of the framework, and the band at 1610 $cm^{-1}$ appears for the glass samples. Another band that appears more clearly in the glass samples from PSP compared to ATR-FTIR is the band at 1330 $cm^{-1}$, corresponding to the in-plane C-N ring stretching and could be more noticeable due to the presence of imidazolate traces, as discussed in section 3.4. More importantly, observing the fingerprint vibrational band at 1099 $cm^{-1}$, we observe a shift in the band towards a lower wavenumber at 1096 $cm^{-1}$ for both glass samples. This is explained by the internal strains created by melting and rapid quenching of the amorphized materials to form the glass. The shift to a lower wavenumber generally suggests strains associated with material shrinkage from fast cooling due to quenching. Furthermore, Figure 5e shows that the vibrational mode at around 1098 $cm^{-1}$ is broadened in both glass samples with a full width half maximum (FWHM) value of 17.8 and 16.5 $cm^{-1}$ respectively for $Cu(Im)_2$-310/1 and $Cu(Im)_2$-245/10, compared with FWHM of 13.4 $cm^{-1}$ for the $Cu(Im)_2$ crystals. This spectral broadening detected at the nanoscale is mainly due to amorphization of the framework.

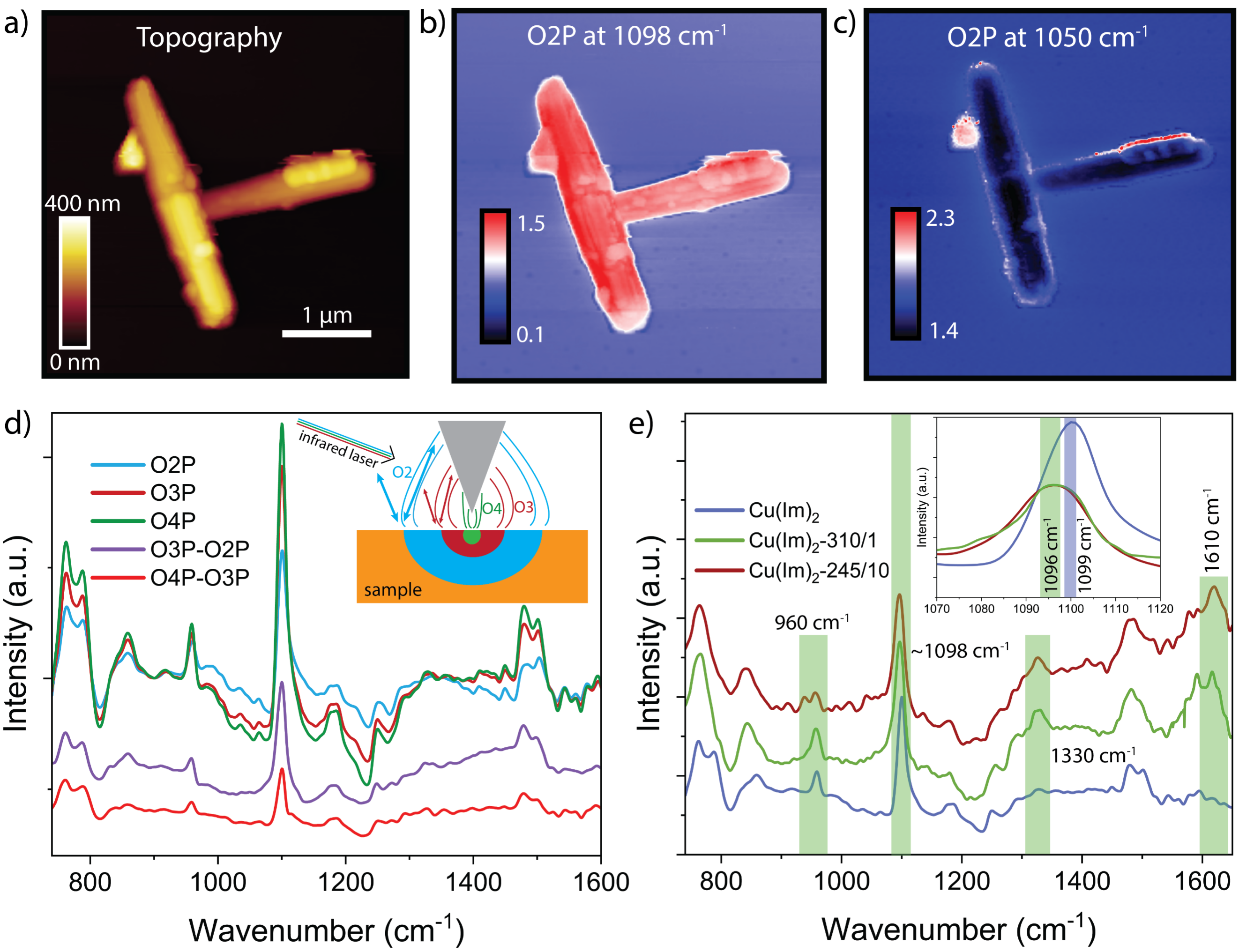

Figure 5. (a) AFM height topography of two overlapping crystals of $Cu(Im)_2$. (b) s-SNOM PsHet nanoimaging of $Cu(Im)_2$ crystals at 1098 $cm^{-1}$ using O2P, showing the crystals in red for absorbing

infrared (IR) signal. (c) PsHet imaging of $Cu(Im)_2$ crystals at 1050 $cm^{-1}$ using O2P, showing the crystals in blue for non-absorbing IR. (d) s-SNOM point nanospectroscopy performed on the $Cu(Im)_2$ crystals from the second, third, and fourth optical harmonic nearfield phase signals (O2P, O3P, O4P), showing the effect of the nearfield signal on the spectra. The difference of the nearfield signals between the consecutive optical harmonics allows the elimination of scattering, reflections and background effects. The inset illustrates the nearfield proximity of each harmonic and its influence on the spectra from the surrounding environment. (e) Point nanospectroscopy performed on the $Cu(Im)_2$ crystals and $Cu(Im)_2$-310/1 and $Cu(Im)_2$-245/10, showing the fourth harmonic optical phase signal O4P. The inset shows the peak shift in the glass samples to a lower wavenumber.

### 3.7 Nanoindentation and Fracture Toughness Studies

For nanoindentation characterization of the glasses, multiple samples of $Cu(Im)_2$ were mounted in an epoxy substrate and finely polished to a surface finish of ~1 µm. This preparation step is to ensure that the surface scratches and uneven features on the glass, which may influence surface detection and accuracy of measurements for indentation experiments, are minimal. The single crystals are too small in size, so our indentation study was limited to glasses of $Cu(Im)_2$-310/1 and $Cu(Im)_2$-245/10. The indentation load versus surface penetration depth ($P-h$) curves and the residual indents are shown in Figure 6a. The load-depth curves show consistency in measurements and high repeatability for each sample, and this is depicted in the small standard deviation of the results. The indentation modulus ($M$) and hardness ($H$) values measured by continuous stiffness measurement (CSM) method is presented in Figure 6b, and their values were averaged between an indentation depth of 500 nm and 2000 nm, derived from a total of 25 indents per glass sample. The results show an indentation modulus $M$ = 11.29 ± 0.24 GPa and hardness $H$ = 833 ± 38 MPa for $Cu(Im)_2$-310/1, and an indentation modulus $M$ = 9.67 ± 0.21 GPa, and hardness $H$ = 819 ± 41 MPa for $Cu(Im)_2$-245/10. Although the results are of a similar order of magnitude, the lower stiffness and hardness of the $Cu(Im)_2$-245/10 can be explained by the prolonged heating of the material leading to framework decomposition, which is consistent with the foregoing characterization. The values of the indentation modulus are close to the highest reported values measured for MOF structures with a low porosity, such as ZIF-zni (Young's modulus, $E_{max}$ ~ 10 GPa) [60], and are surpassing the stiffness and hardness values of ZIF melt-quenched glasses reported to date, namely $a_g$ZIF-4, $a_g$ZIF-62, $a_g$ZIF-76, and $a_g$ZIF-76-mbIm ($E$ ~ 6-8 GPa; $H$ ~ 650-680 MPa) [35, 65].

The subject of fracture being the measure of a material's resistance to cracking (as characterized by fracture toughness, $K_{Ic}$) on MOFs and MOF glasses has been little studied. Knowledge of fracture toughness is important for utilizing MOFs and processing them into larger scale practical applications that require robust and tough structures. Reported toughness values for MOF materials usually employ a constant strain rate and a cube-corner indenter to yield radial cracking [60]. For the study of the fracture toughness of $Cu(Im)_2$ glasses, we employed an indentation-based approach with a cube-corner indenter, a constant strain rate of 0.2 $s^{-1}$, and a maximum load of 50 mN. We estimated the fracture toughness using the Laugier's empirical formula [66] utilizing data derived from indentation cracking experiments:

$$K_{\mathrm{Ic}} = k\left(\frac{a}{l}\right)^{1/2}\left(\frac{E}{H}\right)^{2/3}\frac{P_{\mathrm{m}}}{c^{3/2}}$$

where $k$ is a calibration constant dependent on the indenter geometry, $a$, $l$, and $c$ are dimensions from indentation with $a$ being the length of the residual indent from the center, and $l$ is the crack length. $E$ (corrected from $M$ using Poisson's ratio, assuming $v$ = 0.4 [67]; see SI section 4), $H$, and $P_m$ are the Young's modulus, hardness, and maximum indentation load, respectively. The subscript I denotes the mode-I crack opening.

Figure 6c shows the residual indents and cracks observed for $Cu(Im)_2$-310/1 and $Cu(Im)_2$-245/10 for 1 second and 10 seconds of indenter holding time, respectively. The SEM micrographs show that both radial cracks and lateral cracks formed in some of the samples. The lateral cracks are observed for the $Cu(Im)_2$-245/10 glass sample probably due to the further decomposed structure. For the $Cu(Im)_2$-310/1 glass, shear bands are observed around the indents suggesting a similar fracture mechanism with more resistance to lateral cracking. Both phenomena have been observed before on MOF materials and glasses [43, 48]. This is quantitatively determined employing Laugier's formula, by measuring the radial cracks depicted in Figure 6c. We established that $Cu(Im)_2$-310/1 has a fracture toughness of around 1.5 times higher than $Cu(Im)_2$-245/10, see Figure 6d. Moreover, it was found that the longer holding time causes a slight reduction in the fracture toughness of both samples, although their average values lie within the error bars of the holding time of 1 second. This is likely due to the brittle nature of the glass, where a longer exposure to indentation load causes crack propagation over plastic deformation. In contrast, this finding is different from polycrystalline MOF monoliths where a longer holding time favors shorter cracks due to energy dissipation *via* plastic deformation [49]. Furthermore, the higher error bars in the experiments with a holding time of 1 second are generally higher than the experiments at a holding time of 10 seconds. This is explained as the short indentation duration and the quick withdrawal of the indenter causes a fast impact on the material, thereby resulting in more cracking especially by allowing the lateral cracks to propagate. The values of the measured fracture toughness for both samples are tabulated in Table S2.

The fracture toughness of both glass samples is remarkably high and to our knowledge is at least three times higher than any previously reported fracture toughness of MOF glasses lying around 0.1 MPa $m^{1/2}$ [60], see $K_{Ic}$ data in Table S3. For instance, some of us recently reported a flux-mediated ligand exchange ZIF glass, termed $gphen_{0.49}ZIF-62(Co)$ that exhibits $K_{Ic}$ = 0.103 ± 0.041 MPa $m^{1/2}$ [40]. Interestingly, $Cu(Im)_2$-310/1 has shown a notably high toughness, placing it just below HKUST-1 among what is reported so far on the fracture toughness of MOFs, not limited to just glasses [48, 60]. This places $Cu(Im)_2$ glass as a promising candidate for future applications that require shaping of MOFs while maintaining mechanical resilience. The advantage of $Cu(Im)_2$ glass over HKUST-1 is that $Cu(Im)_2$ is meltable and thus can potentially be shaped into thin films. Additionally, it can host other MOF nanocrystals to yield porosity or tune the functionality of the resulting films given the relatively lower melting temperature of the $Cu(Im)_2$ crystals compared to other glass forming MOFs [37].

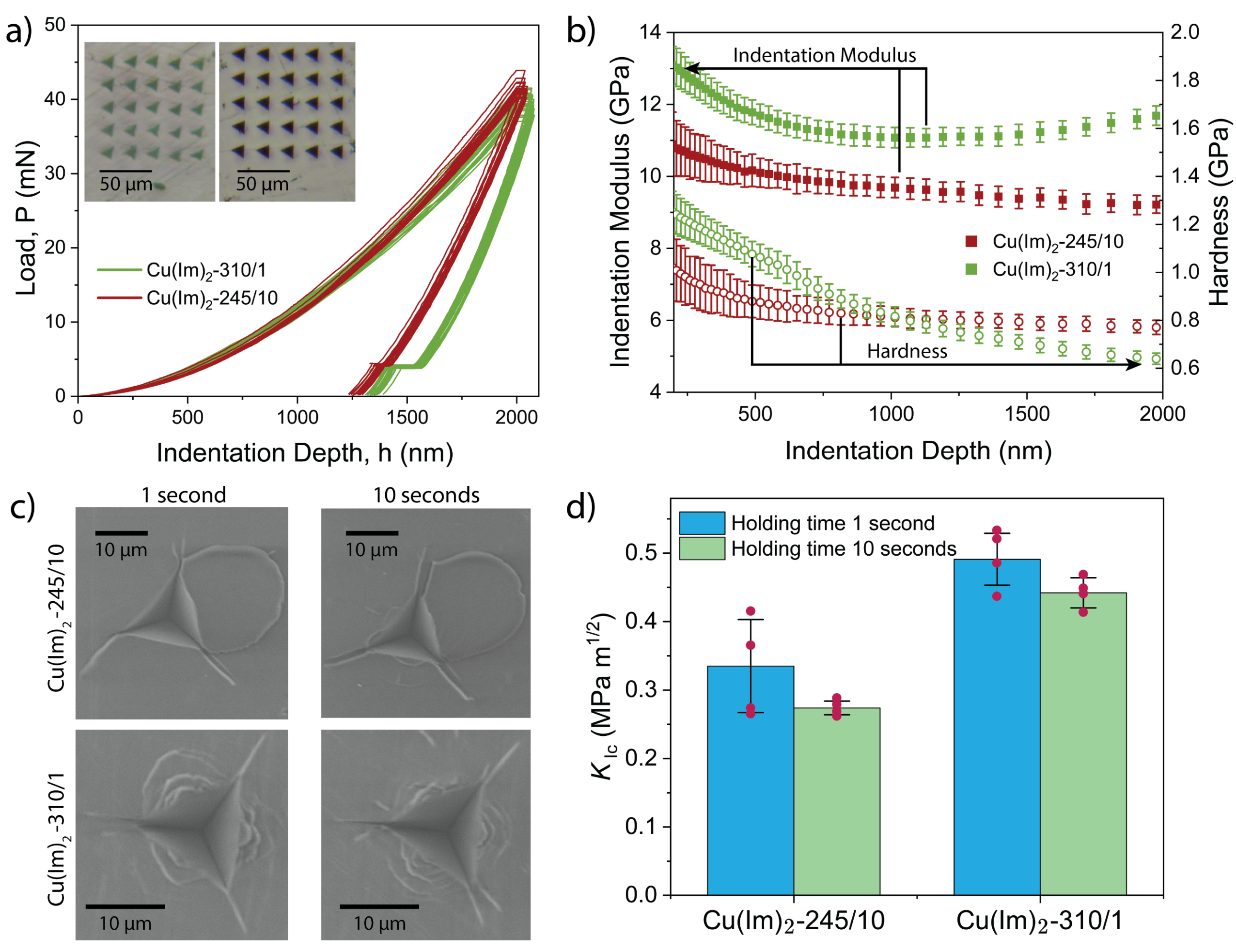

Figure 6. (a) Indentation load-depth ($P-h$) curves for the glasses of $Cu(Im)_2$-310/1 and $Cu(Im)_2$-245/10. The horizontal test segment in the unloading curve (at 10 % maximum load) was used for thermal drift correction. The residual indents are shown for the plotted data of $Cu(Im)_2$-310/1 (left) and $Cu(Im)_2$-245/10 (right). (b) Indentation modulus (in full squares) and hardness (in empty circles) as a function of indentation depth determined by CSM method for the glasses of $Cu(Im)_2$-310/1 (green) and $Cu(Im)_2$-245/10 (red). (c) SEM micrographs showing residual indents and cracks for $Cu(Im)_2$-310/1 and $Cu(Im)_2$-245/10 tested under a maximum load of 50 mN for 1 second and 10 seconds, respectively. (d) Bar charts of the fracture toughness at different holding times for $Cu(Im)_2$-310/1 and $Cu(Im)_2$-245/10. The error bars were derived from 4 indentation measurements.

## 4 Conclusions

In this study, we presented a crystalline copper-based metal-organic framework, $Cu(Im)_2$, which was synthesized using a sol-gel process and subsequently melt-quenched into a glass. We showed that the resulting glass is not a decomposed material rather a reduced form that reorganizes its structure likely to a Cu(Im) structure, and we characterized the glass along with the original MOF nanocrystals through various multiscale and multimodal characterization techniques. We showed that melting the $Cu(Im)_2$ crystals at a high temperature for a shorter time ($Cu(Im)_2$-310/1) produces a glass with the least partial chemical decomposition and improved stiffness and toughness. The $Cu(Im)_2$ glasses have shown a high indentation modulus and hardness combined with an exceptionally high fracture toughness, surpassing the toughest MOF glass reported to our knowledge. This advancement in fabricating MOF glass at a reasonably low temperature (~300 °C) combined with high stiffness and toughness bodes well for future work in the shaping of MOF glasses to realize applications requiring fracture resistant structures. The fact that the $Cu(Im)_2$ crystals melt at a temperature below the decomposition temperature of many MOF nanocrystals allows this material to host non-meltable MOFs with tunable functional properties, generating bespoke geometry for a specific application together with mechanical robustness.

**Acknowledgements**

M.E.S. acknowledges the Oxford-Qatar-Thatcher Graduate Scholarship for DPhil postgraduate funding. J.C.T. thanks the UKRI Engineering and Physical Sciences Research Council (EPSRC) award (TEGMOF EP/Z534146/1) for equipment funding. We thank Diamond Light Source for the award of beamtimes SM40142 and SM36374 on beamline B22 MIRIAM, and we thank the assistance from Hendrik Vondracek and Alexander Hawkins on the beamline.

## References

1. O'Keeffe, M., et al., *Frameworks for extended solids: Geometrical design principles*. J. Solid State Chem., 2000. **152**: 3-20.
2. Yaghi, O.M., et al., *Reticular synthesis and the design of new materials*. Nature, 2003. **423**: 705-714.
3. Tian, T., et al., *A sol–gel monolithic metal–organic framework with enhanced methane uptake*. Nat. Mater., 2018. **17**: 174-179.
4. Tricarico, M., et al., *Exploring the Photophysical and Mechanical Behavior of Fluorescent Metal-Organic Framework Monoliths*. Chem. Mater., 2024. **36**: 8247-8254.
5. Gutiérrez, M., Y. Zhang, and J.C. Tan, *Confinement of Luminescent Guests in Metal-Organic Frameworks: Understanding Pathways from Synthesis and Multimodal Characterization to Potential Applications of LG@MOF Systems*. Chem. Rev., 2022. **122**: 10438-10483.
6. Allendorf, M.D., et al., *Guest-Induced Emergent Properties in Metal-Organic Frameworks*. J. Phys. Chem. Lett., 2015. **6**: 1182-1195.
7. Liu, X.T., et al., *Recent progress in host-guest metal-organic frameworks: Construction and emergent properties*. Coord. Chem. Rev., 2023. **476**: 214921.
8. Slater, B. and J.C. Tan, *Triboelectric behaviour of selected zeolitic-imidazolate frameworks: exploring chemical, morphological and topological influences*. Chem. Sci., 2024. **15**: 10056-10064.
9. Ye, J.H. and J.C. Tan, *High-performance triboelectric nanogenerators incorporating chlorinated zeolitic imidazolate frameworks with topologically tunable dielectric and surface adhesion properties*. Nano Energy, 2023. **114**: 108687.
10. Mondloch, J.E., et al., *Destruction of chemical warfare agents using metal-organic frameworks*. Nat. Mater., 2015. **14**: 512-516.
11. Son, F., et al., *Uncovering the Role of Metal-Organic Framework Topology on the Capture and Reactivity of Chemical Warfare Agents*. Chem. Mater., 2020. **32**: 4609-4617.
12. Song, W., et al., *MOF water harvester produces water from Death Valley desert air in ambient sunlight*. Nature Water, 2023. **1**: 626–634.
13. *A market for metal-organic frameworks*. Nat. Mater., 2025. **24**: 157-157.
14. Stassen, I., et al., *Electrochemical Film Deposition of the Zirconium Metal-Organic Framework UiO-66 and Application in a Miniaturized Sorbent Trap*. Chem. Mater., 2015. **27**: 1801-1807.
15. Buchan, I., M.R. Ryder, and J.C. Tan, *Micromechanical Behavior of Polycrystalline Metal-Organic Framework Thin Films Synthesized by Electrochemical Reaction*. Cryst. Growth Des., 2015. **15**: 1991-1999.
16. Carrington, M.E., et al., *Sol-gel processing of a covalent organic framework for the generation of hierarchically porous monolithic adsorbents*. Chem., 2022. **8**: 2961-2977.
17. Hou, J.W., A.F. Sapnik, and T.D. Bennett, *Metal-organic framework gels and monoliths*. Chem. Sci., 2020. **11**: 310-323.
18. Olajire, A.A., *Synthesis chemistry of metal-organic frameworks for CO capture and conversion for sustainable energy future*. Renewable & Sustainable Energy Reviews, 2018. **92**: 570-607.
19. Fonseca, J. and T.H. Gong, *Fabrication of metal-organic framework architectures with macroscopic size: A review*. Coord. Chem. Rev., 2022. **462**: 214520.
20. Kalaj, M., et al., *MOF-Polymer Hybrid Materials: From Simple Composites to Tailored Architectures*. Chem. Rev., 2020. **120**: 8267-8302.
21. Mahdi, E.M. and J.C. Tan, *Mixed-matrix membranes of zeolitic imidazolate framework (ZIF-8)/Matrimid nanocomposite: Thermo-mechanical stability and viscoelasticity underpinning membrane separation performance*. J. Membr. Sci., 2016. **498**: 276-290.
22. Chen, Y.F., et al., *Shaping of Metal-Organic Frameworks: From Fluid to Shaped Bodies and Robust Foams*. J. Am. Chem. Soc., 2016. **138**: 10810-10813.

23. Thakkar, H., et al., *3D-Printed Metal-Organic Framework Monoliths for Gas Adsorption Processes*. ACS Appl. Mater. Interfaces, 2017. **9**: 35908-35916.
24. Claessens, B., et al., *3D-Printed ZIF-8 Monoliths for Biobutanol Recovery*. Ind. Eng. Chem. Res., 2020. **59**: 8813-8824.
25. Katayama, Y., K.C. Bentz, and S.M. Cohen, *Defect-Free MOF-Based Mixed-Matrix Membranes Obtained by Corona Cross-Linking*. ACS Appl. Mater. Interfaces, 2019. **11**: 13029-13037.
26. Zhao, Z.J., et al., *Preparation of ZIF-62 polycrystalline and glass membranes for helium separation*. J. Membr. Sci., 2024. **700**: 122677.
27. Li, J., et al., *Coordination Polymer Glasses with Lava and Healing Ability for High-Performance Gas Sieving*. Angew. Chem. Int. Ed., 2021. **60**: 21304-21309.
28. Stone, D.M., et al., *Control of ZIF-62 and aZIF-62 Film Thickness within Asymmetric Tubular Supports through Pressure and Dose Time Variation of Atomic Layer Deposition*. Small, 2024. **20**: 2307202.
29. Smirnova, O., et al., *Micro-optical elements from optical-quality ZIF-62 hybrid glasses by hot imprinting*. Nat. Commun., 2024. **15**: 5079.
30. Xue, W.L., et al., *Insights Into the Mechanochemical Glass Formation of Zeolitic Imidazolate Frameworks*. Angew. Chem. Int. Ed., 2024. **63**: e202405307.
31. León-Alcaide, L., et al., *Meltable, Glass-Forming, Iron Zeolitic Imidazolate Frameworks*. J. Am. Chem. Soc., 2023. **145**: 11258-11264.
32. Xue, W.-L., et al., *Mechanochemical Synthesis Enables Melting, Glass Formation and Glass–Ceramic Conversion in a Cadmium-Based Zeolitic Imidazolate Framework*. J. Am. Chem. Soc., 2025. **147**: 15625-15635.
33. Bennett, T.D., et al., *Thermal Amorphization of Zeolitic Imidazolate Frameworks*. Angew. Chem. Int. Ed., 2011. **50**: 3067-3071.
34. Wei, Y.S., et al., *Hierarchical Metal-Organic Network-Forming Glasses toward Applications*. Adv. Funct. Mater., 2023. **34**: 2307226.
35. Bennett, T.D., et al., *Melt-Quenched Glasses of Metal-Organic Frameworks*. J. Am. Chem. Soc., 2016. **138**: 3484-3492.
36. Bumstead, A.M., et al., *Investigating the melting behaviour of polymorphic zeolitic imidazolate frameworks*. Crystengcomm, 2020. **22**: 3627-3637.
37. Frentzel-Beyme, L., et al., *Quantification of gas-accessible microporosity in metal-organic framework glasses*. Nat. Commun., 2022. **13**: 7750.
38. Bennett, T.D., et al., *Hybrid glasses from strong and fragile metal-organic framework liquids*. Nat. Commun., 2015. **6**: 8079.
39. Nozari, V., et al., *Ionic liquid facilitated melting of the metal-organic framework ZIF-8*. Nat. Commun., 2021. **12**: 5703.
40. Weiß, J.-B., et al., *Flux-mediated ligand exchange restructures metal–organic framework glasses*. Nat. Mater., 2026: Article in Press.
41. Xiong, M., et al., *Unraveling the effects of linker substitution on structural, electronic and optical properties of amorphous zeolitic imidazolate frameworks-62 (a-ZIF-62) glasses: a DFT study (vol 10, pg 14013, 2020)*. RSC Adv., 2020. **10**: 16804-16804.
42. Xu, X.Y., T. Du, and M.M. Smedskjaer, *Mechanical properties of zeolitic imidazolate framework crystal-glass composites: A molecular dynamics study*. Journal of Non-Crystalline Solids, 2025. **650**: 123379.
43. Stepniewska, M., et al., *Observation of indentation-induced shear bands in a metal-organic framework glass*. Proc. Natl. Acad. Sci. USA, 2020. **117**: 10149-10154.
44. Li, S., et al., *Mechanical Properties and Processing Techniques of Bulk Metal-Organic Framework Glasses*. J. Am. Chem. Soc., 2019. **141**: 1027-1034.
45. Widmer, R.N., et al., *Plasticity of Metal-Organic Framework Glasses*. J. Am. Chem. Soc., 2021. **143**: 20717-20724.

46. To, T., et al., *Fracture toughness of a metal-organic framework glass.* Nat. Commun., 2020. **11**: 2593.
47. Tan, J.C., et al., *Anisotropic mechanical properties of polymorphic hybrid inorganic-organic framework materials with different dimensionalities.* Acta. Mater., 2009. **57**: 3481-3496.
48. Tricarico, M. and J.-C. Tan, *Nanostructure-dependent indentation fracture toughness of metal-organic framework monoliths.* Next Materials, 2023. **1**: 100009.
49. El Skafi, M., et al., *Mechanical behavior of metal-organic framework monoliths with enhanced fracture toughness.* Mater. Today Nano, 2026. **33**: 100760.
50. Xue, W.L., et al., *Highly porous metal-organic framework liquids and glasses via a solvent-assisted linker exchange strategy of ZIF-8.* Nat. Commun., 2024. **15**: 4420.
51. McHugh, L.N. and T.D. Bennett, *Introducing porosity into metal-organic framework glasses.* J. Mater. Chem. A, 2022. **10**: 19552-19559.
52. Kolodzeiski, P., et al., *Alkali-ion-modified zeolitic imidazolate framework glasses.* Nat. Chem., 2026. **18**: 1383–1392.
53. Li, D.D., et al., *Self-supported flux melted glass membranes fabricated by melt quenching for gas separation.* J. Membr. Sci., 2024. **695**: 122492.
54. Longley, L., et al., *Flux melting of metal-organic frameworks.* Chem. Sci., 2019. **10**: 3592-3601.
55. Tuffnell, J.M., et al., *Novel metal-organic framework materials: blends, liquids, glasses and crystal-glass composites.* Chem. Commun., 2019. **55**: 8705-8715.
56. Watcharatpong, T., et al., *Alloying One-Dimensional Coordination Polymers To Create Ductile Materials.* J. Am. Chem. Soc., 2024. **146**: 23412-23416.
57. Watcharatpong, T., et al., *Coordination polymer-forming liquid Cu(2-isopropylimidazolate).* Chem. Sci., 2022. **13**: 11422-11426.
58. Masciocchi, N., et al., *Extended polymorphism in copper(II) imidazolate polymers: A spectroscopic and XRPD structural study.* Inorg. Chem., 2001. **40**: 5897-5905.
59. Tian, Y.Q., et al., *[Cu(I)(im)]: Is this air-stable copper(I) imidazolate (8^2 10)-net polymer the species responsible for the corrosion-inhibiting properties of imidazole with copper metal?* Eur. J. Inorg. Chem., 2004: 1813-1816.
60. Tan, J.-C., *Mechanical Behaviour of Metal – Organic Framework Materials*, ed. J.-C. Tan. 2023: The Royal Society of Chemistry.
61. Attocube, *AFM-IR vs. s-SNOM: Applications Comparison:* https://www.attocube.com/application/files/1517/0901/8239/AFM-IR_vs._s-SNOM_-_Applications_Comparison.pdf. 2024.
62. Mester, L., A.A. Govyadinov, and R. Hillenbrand, *High-fidelity nano-FTIR spectroscopy by on-pixel normalization of signal harmonics.* Nanophotonics, 2022. **11**: 377-390.
63. Tan, J.C., et al., *Quantum mechanical predictions to elucidate the anisotropic elastic properties of zeolitic imidazolate frameworks: ZIF-4 and ZIF-zni.* Crystengcomm, 2015. **17**: 375-382.
64. Möslein, A.F., et al., *Near-Field Infrared Nanospectroscopy Reveals Guest Confinement in Metal-Organic Framework Single Crystals.* Nano Lett., 2020. **20**: 7446-7454.
65. Kim, M., et al., *Melt-quenched carboxylate metal-organic framework glasses.* Nat. Commun., 2024. **15**: 1174.
66. Laugier, M.T., *New Formula for Indentation Toughness in Ceramics.* J. Mater. Sci. Lett., 1987. **6**: 355-356.
67. Qiao, A., et al., *A metal-organic framework with ultrahigh glass-forming ability.* Science Advances, 2018. **4**: eaao6827.

*Supplementary Information*

*for*

# Multimodal and Multiscale Interrogation of a Mechanically Tough Glass Forming Copper-Based Metal-Organic Framework

Mounir El Skafi[1], Guo-Qiang Li[2], Sophie R. Thomas[3], James D. Taylor[4], Mark Frogley[5], Gianfelice Cinque[5], Marc Pignitter[3], Michael Reithofer[3], Jia-Min Chin[3], Sebastian Henke[2], and Jin-Chong Tan[1*]

[1]Multifunctional Materials and Composites (MMC) Laboratory, Department of Engineering Science, Parks Road, Oxford, OX1 3PJ, University of Oxford, United Kingdom.

[2]Anorganische Chemie, Fakultät für Chemie und Chemische Biologie, Technische Universität Dortmund, Otto-Hahn Straße 6, 44227, Dortmund, Germany.

[3]Institute of Inorganic Chemistry, University of Vienna, Währinger Str. 42, 1090 Vienna, Austria

[4]ISIS Neutron and Muon Source, Science and Technology Facility Council, UKRI, Rutherford Appleton Laboratory, Chilton, Didcot OX11 0QX, United Kingdom.

[5]Diamond Light Source, Harwell Campus, Chilton, Oxford OX11 0DE, United Kingdom.

[*]*Corresponding author*: jin-chong.tan@eng.ox.ac.uk

Table of Contents

## 1. Supplementary Figures

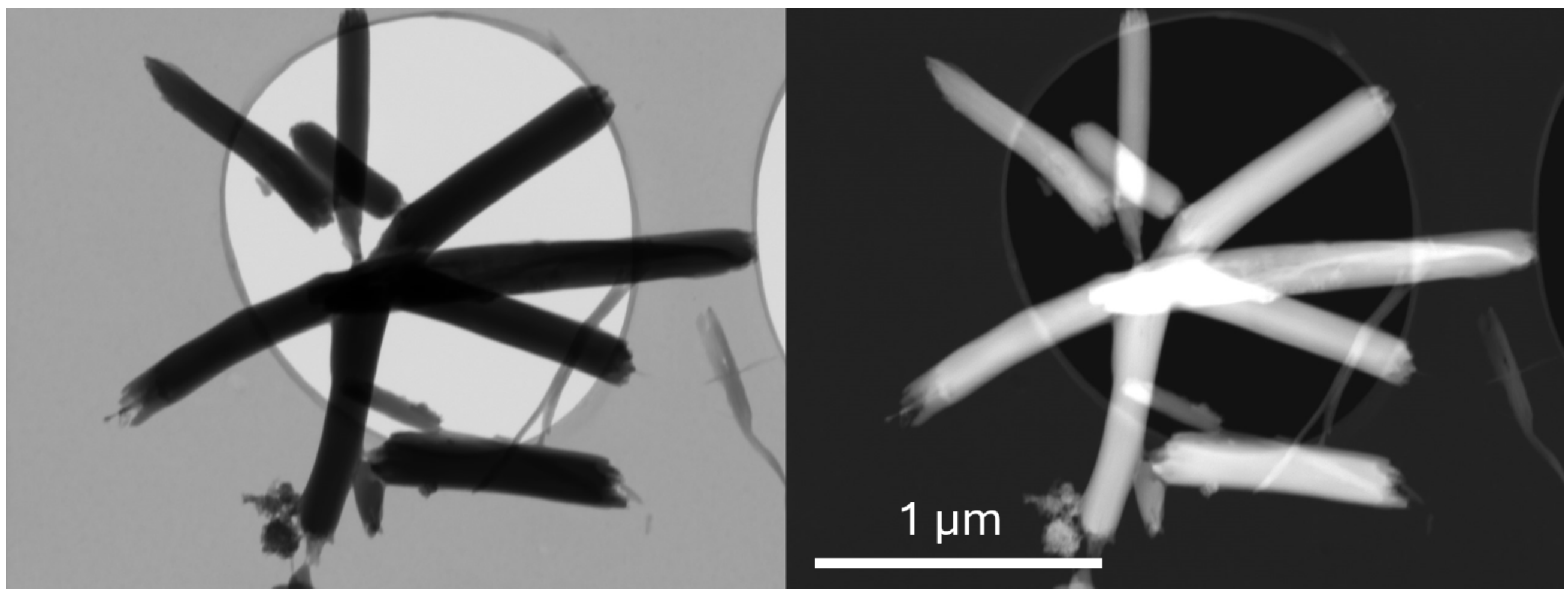

Figure S1. R-STEM micrographs of $Cu(Im)_2$ crystals obtained under the bright mode (left) and the dark mode (right).

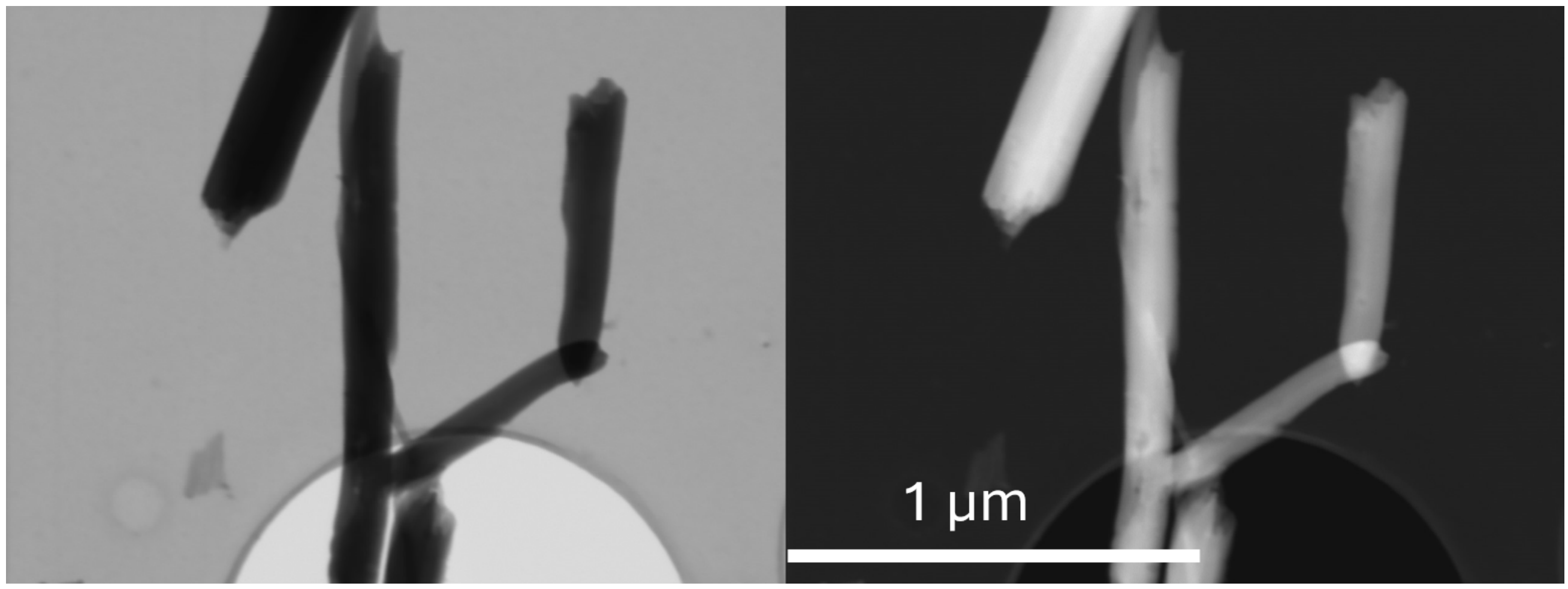

Figure S2. R-STEM micrographs of $Cu(Im)_2$ crystals obtained under the bright mode (left) and the dark mode (right).

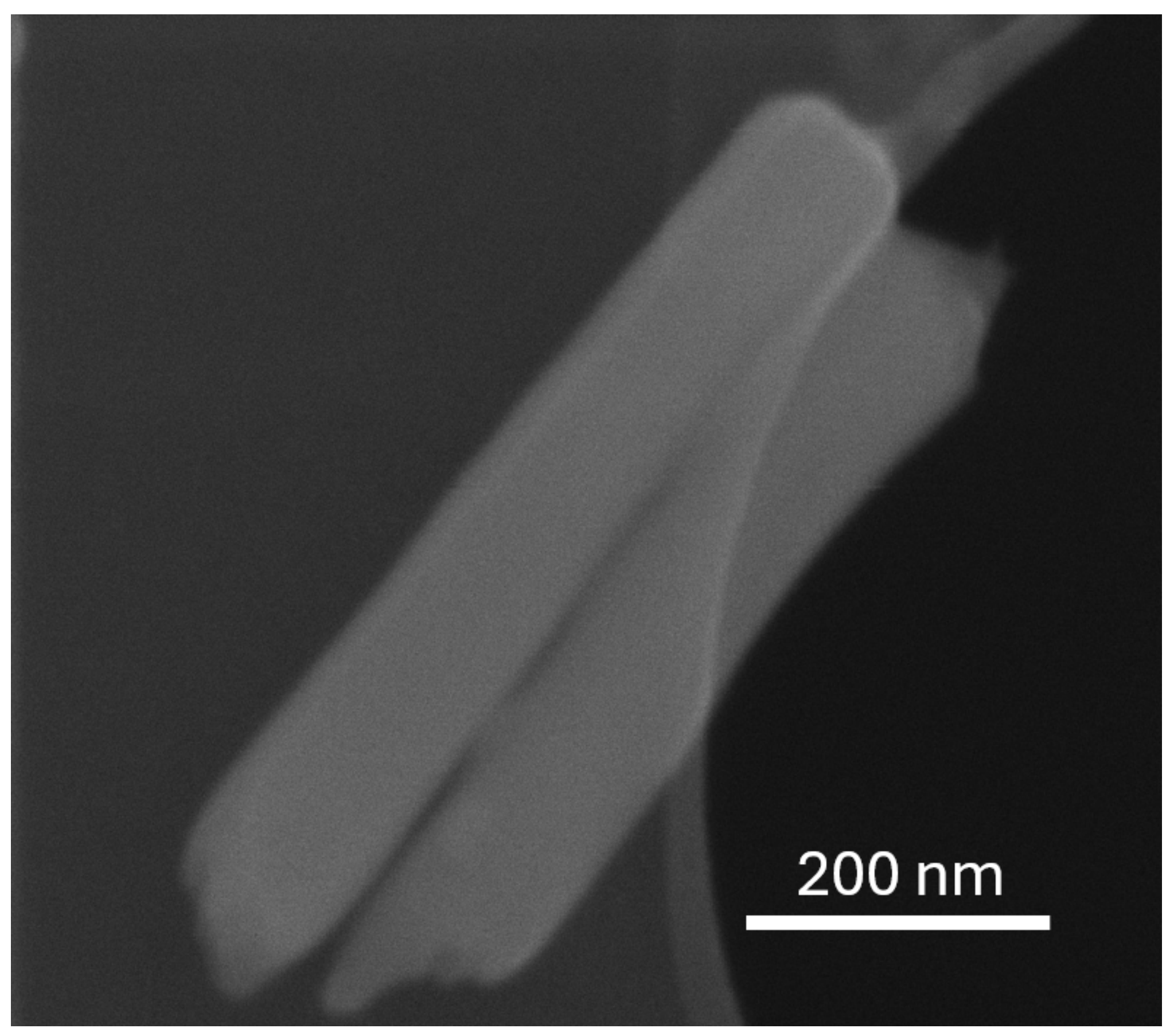

Figure S3. R-STEM micrograph of a pair of $Cu(Im)_2$ crystals in secondary electron (SE) mode.

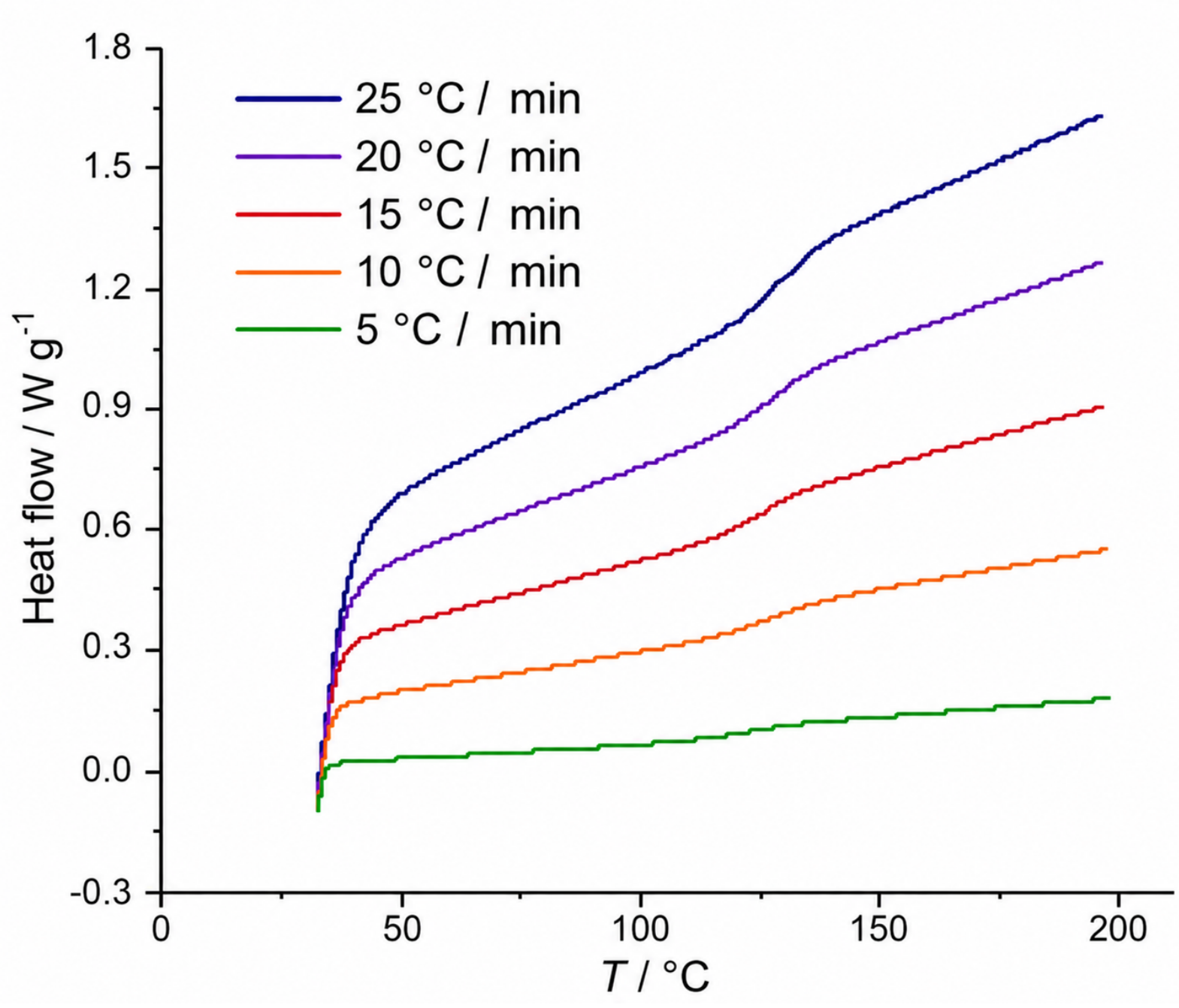

Figure S4. DSC scans at different heating rates, from which it was determined that the glass transition temperature of $Cu(Im)_2$-310/1 glass is at ~110 °C.

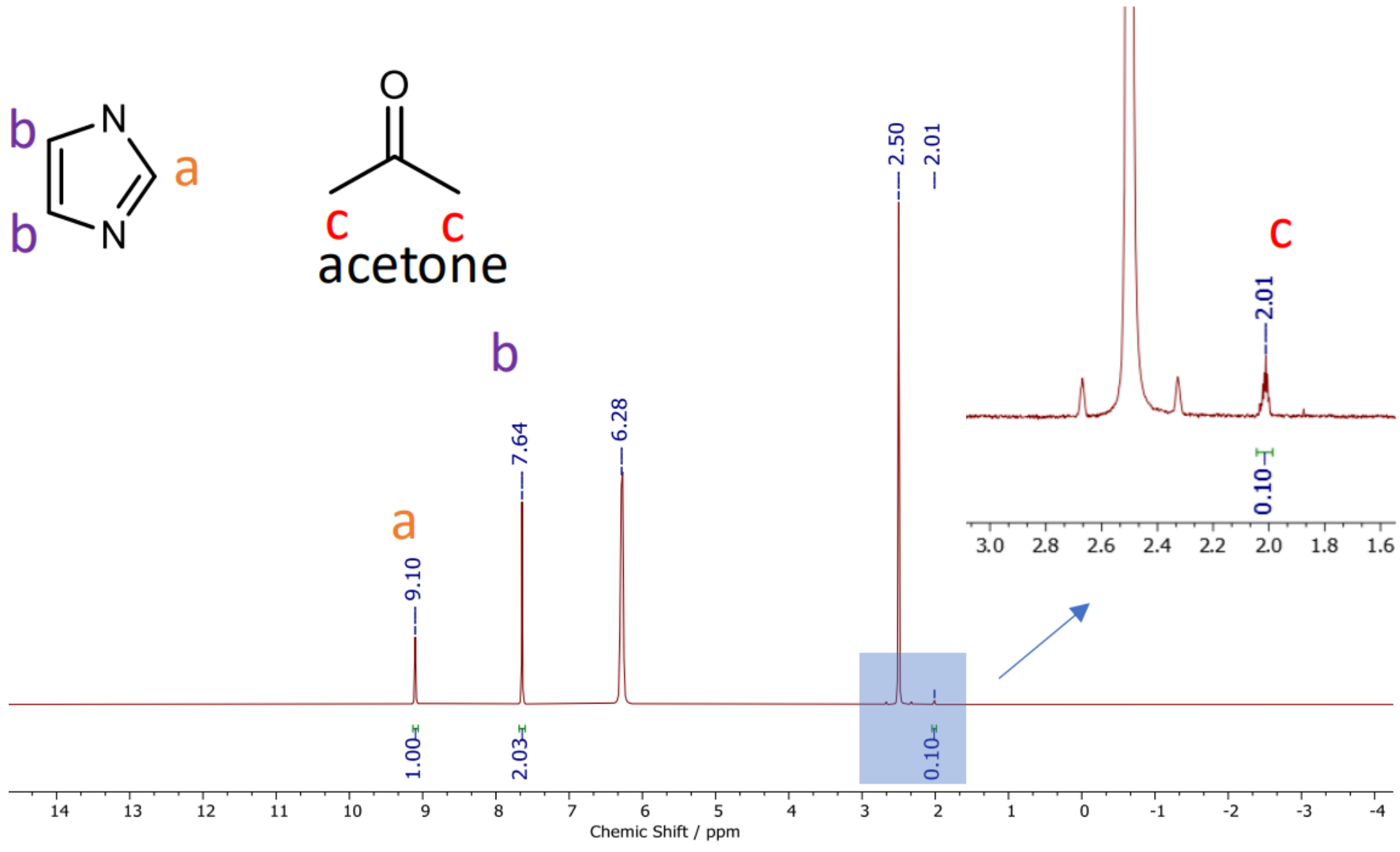

Figure S5. NMR results for $Cu(Im)_2$ crystals in DMSO-$d_6$ + $DCl/D_2O$.

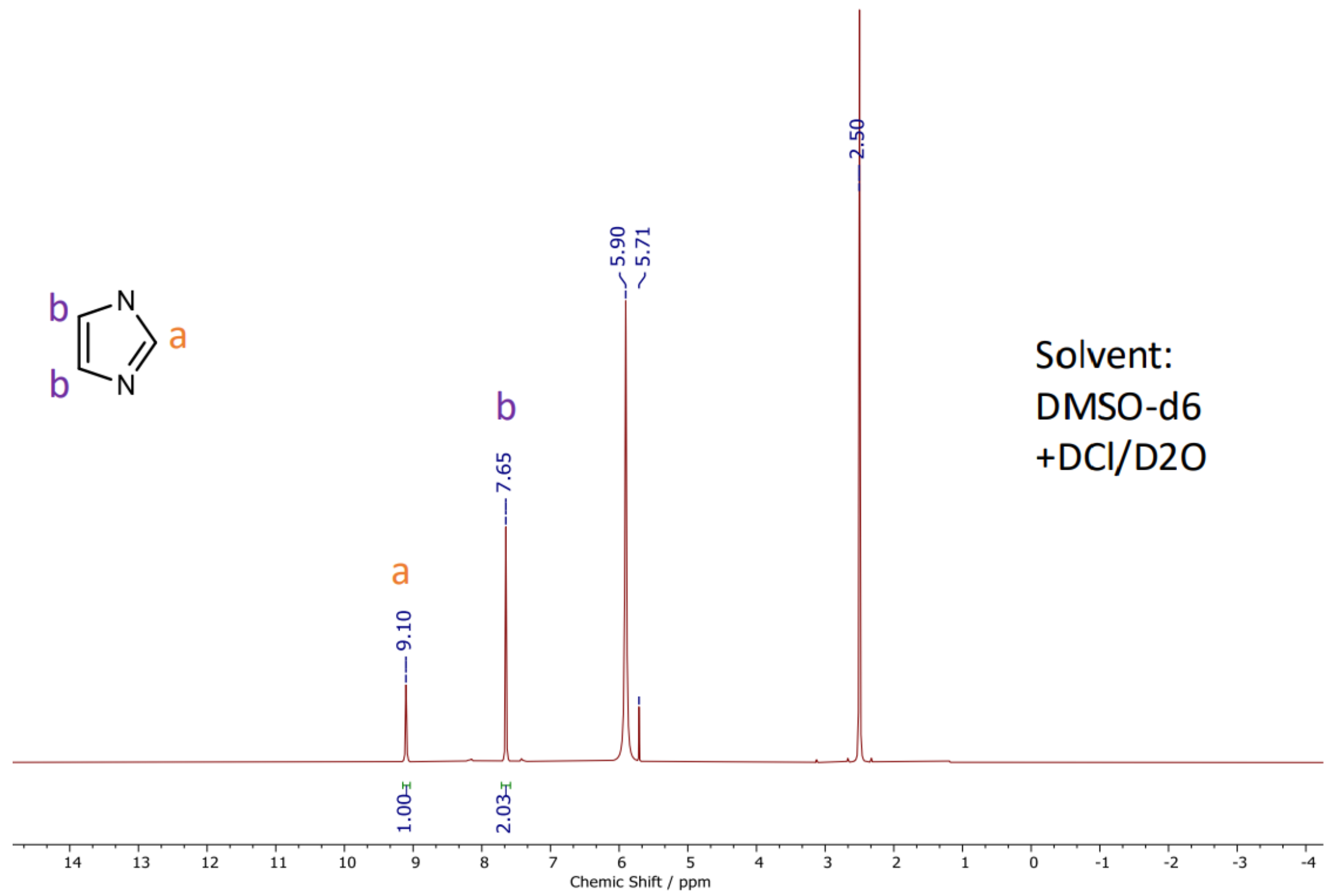

Figure S6. NMR results for $Cu(Im)_2$-240/20 in DMSO-$d_6$ + $DCl/D_2O$.

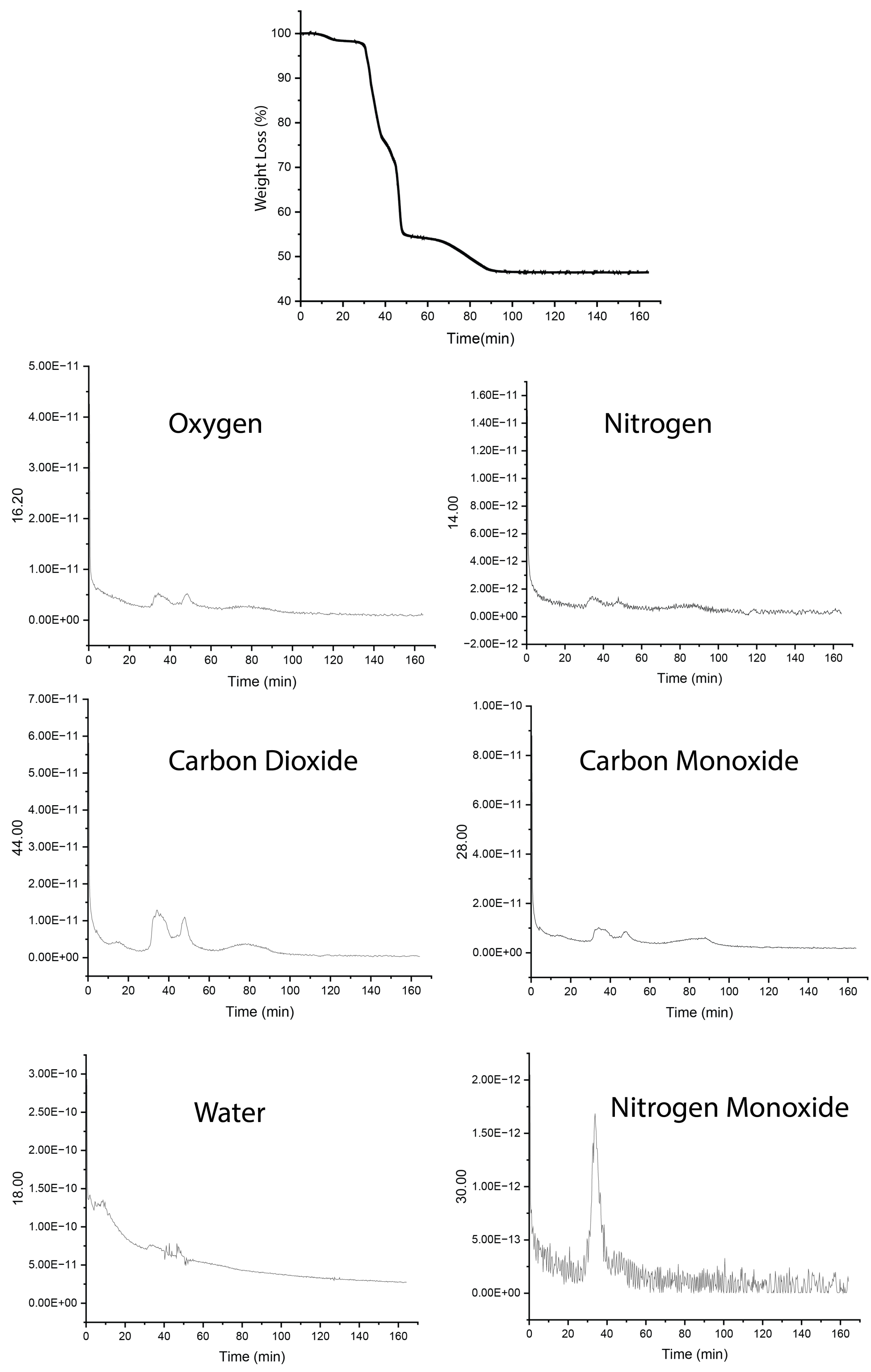

Figure S7. TGA-MS analysis on the $Cu(Im)_2$ crystals showing the species released upon the heating and decomposition of the crystals. The y-axis labels represent the molar masses of the molecules being analyzed. Heating rate is 10 °C/min.

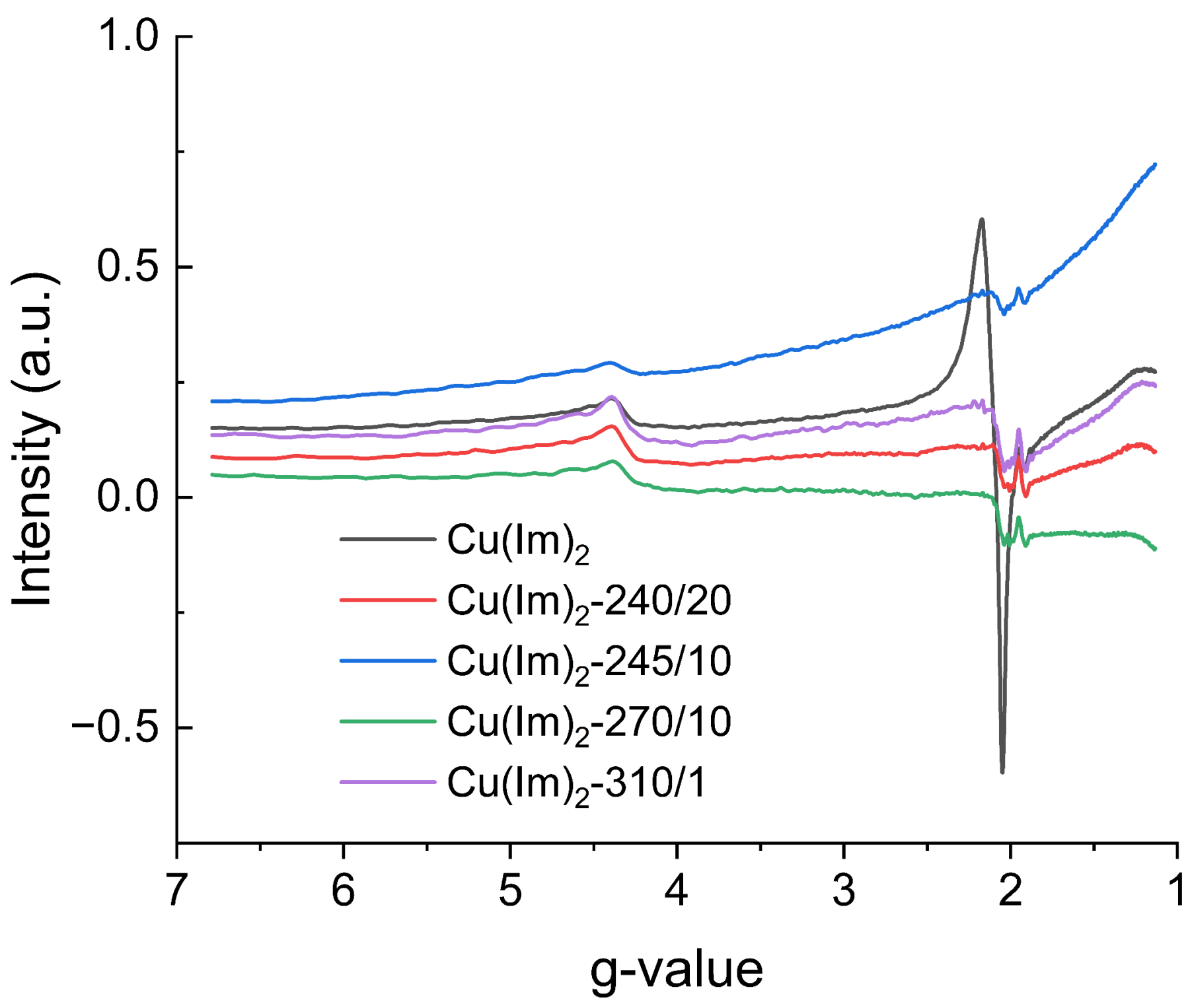

Figure S8. EPR intensity plotted against the g-value for the $Cu(Im)_2$ crystals and glasses.

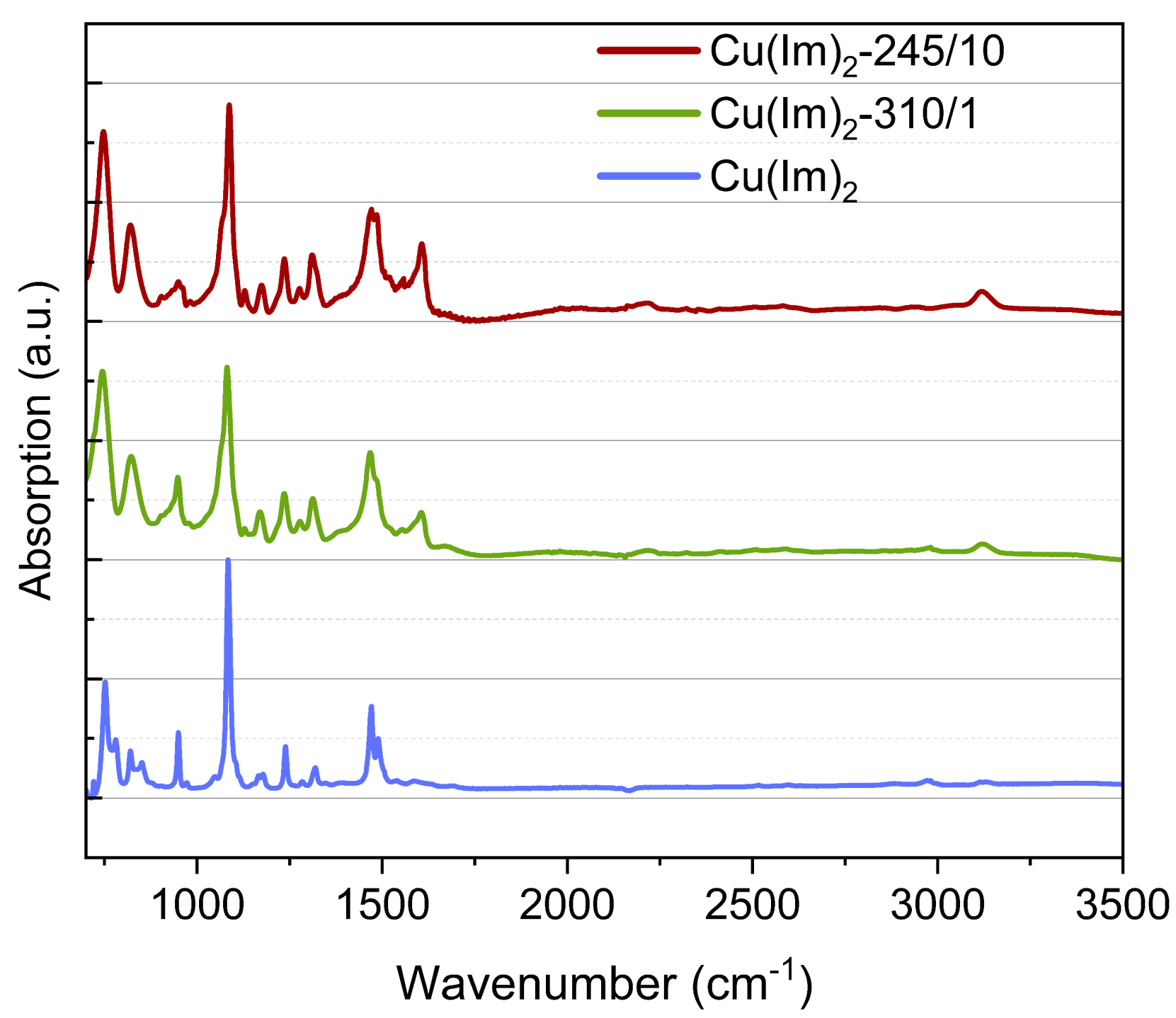

Figure S9. Extended ATR-FTIR spectra of the $Cu(Im)_2$ crystals and glasses prepared under different conditions.

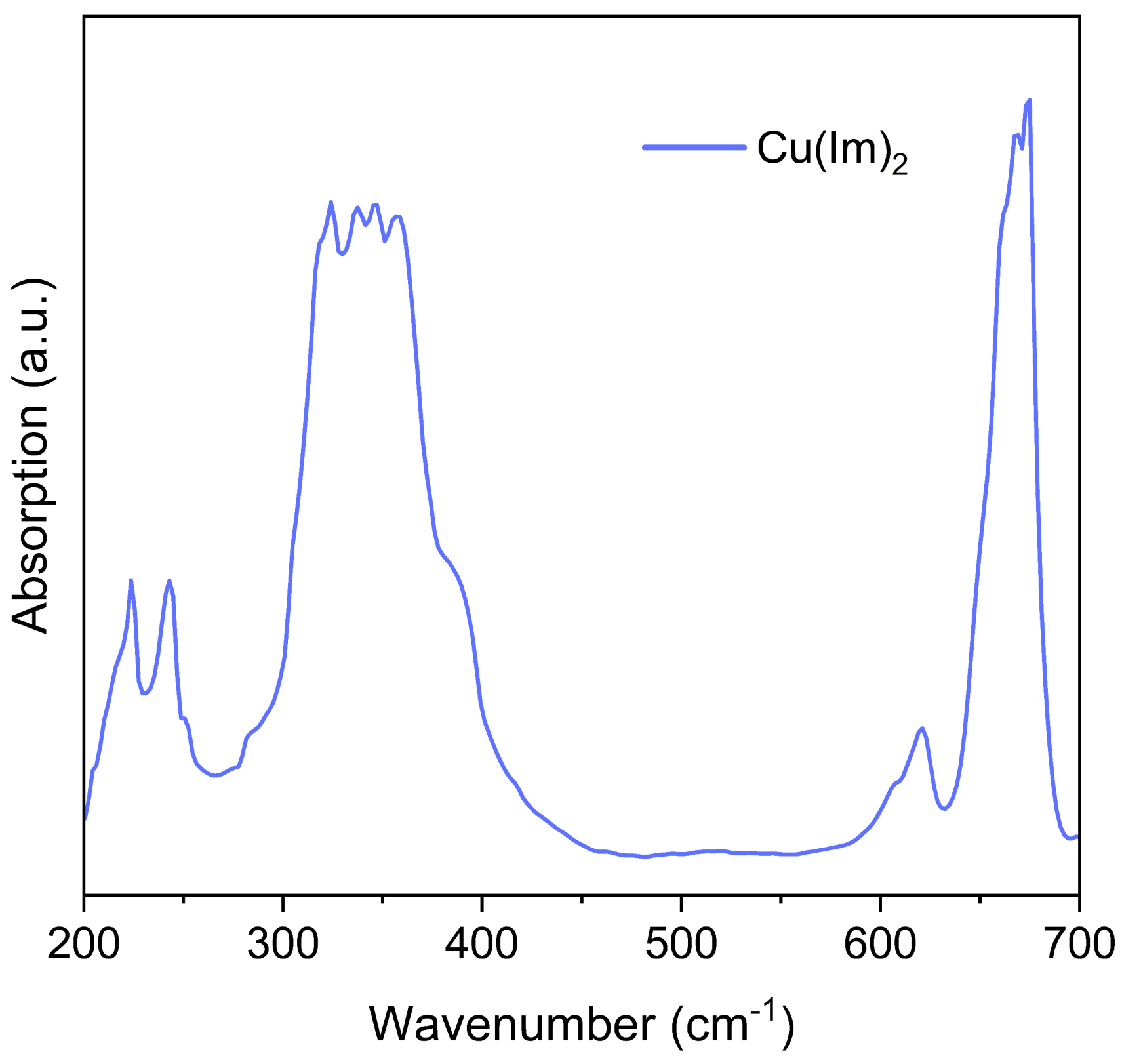

Figure S10. Synchrotron far-IR spectra collected using the IR microscope at Diamond B22.

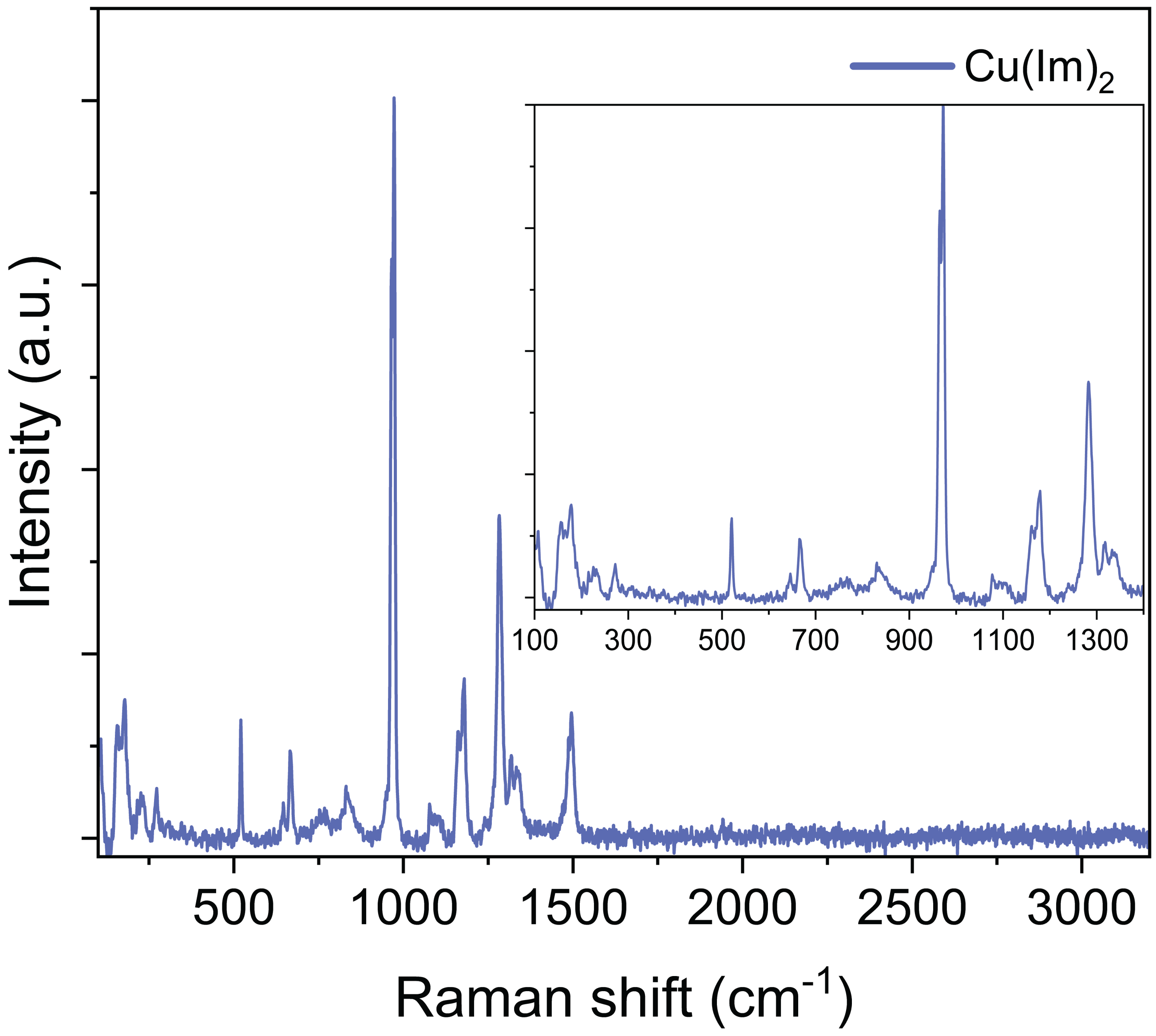

Figure S11. Extended Raman spectra for $Cu(Im)_2$ crystals.

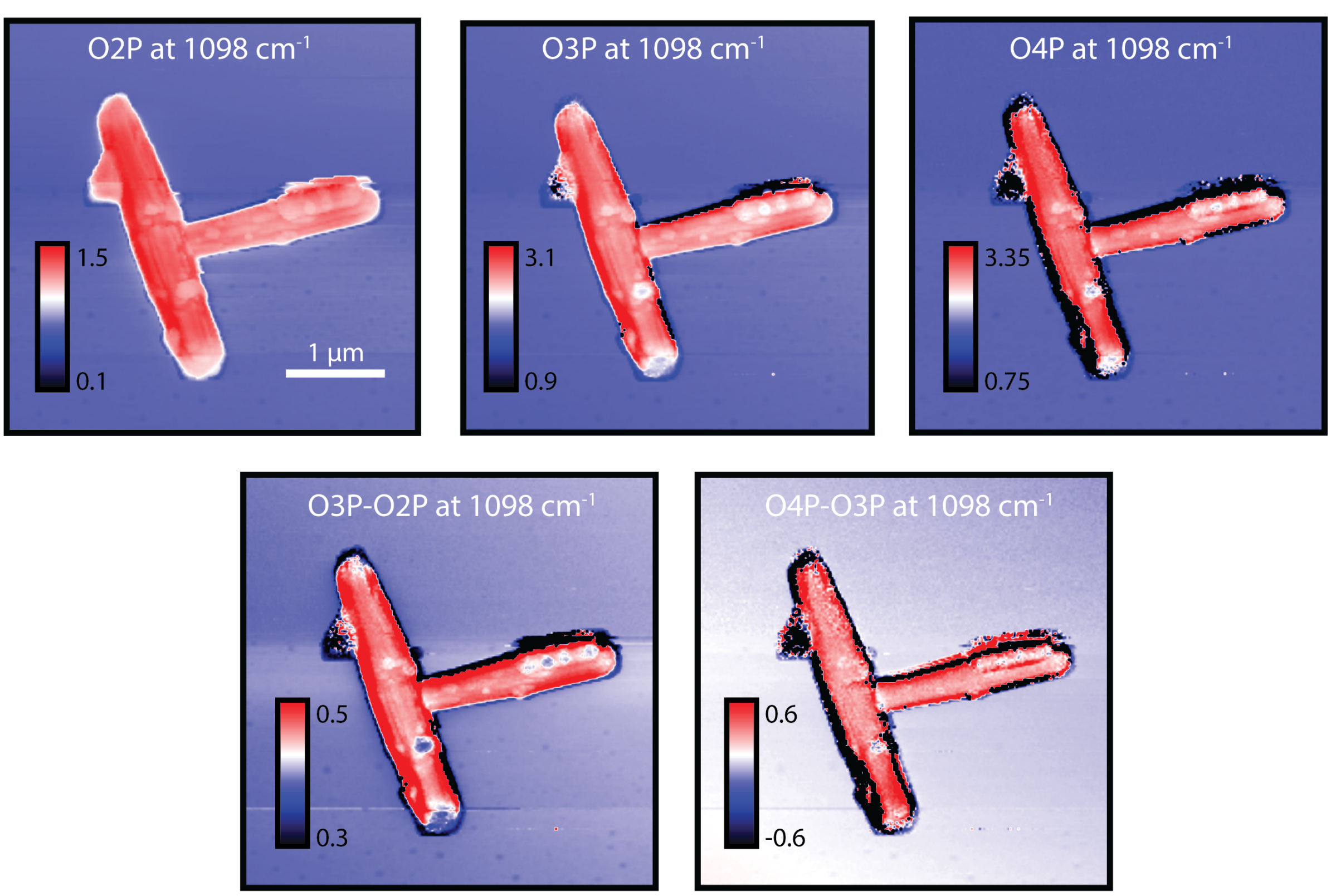

Figure S12. PsHet imaging of $Cu(Im)_2$ crystals measured under different nearfield optical harmonics, namely O2P to O4P and background subtraction from the differential phase signals.

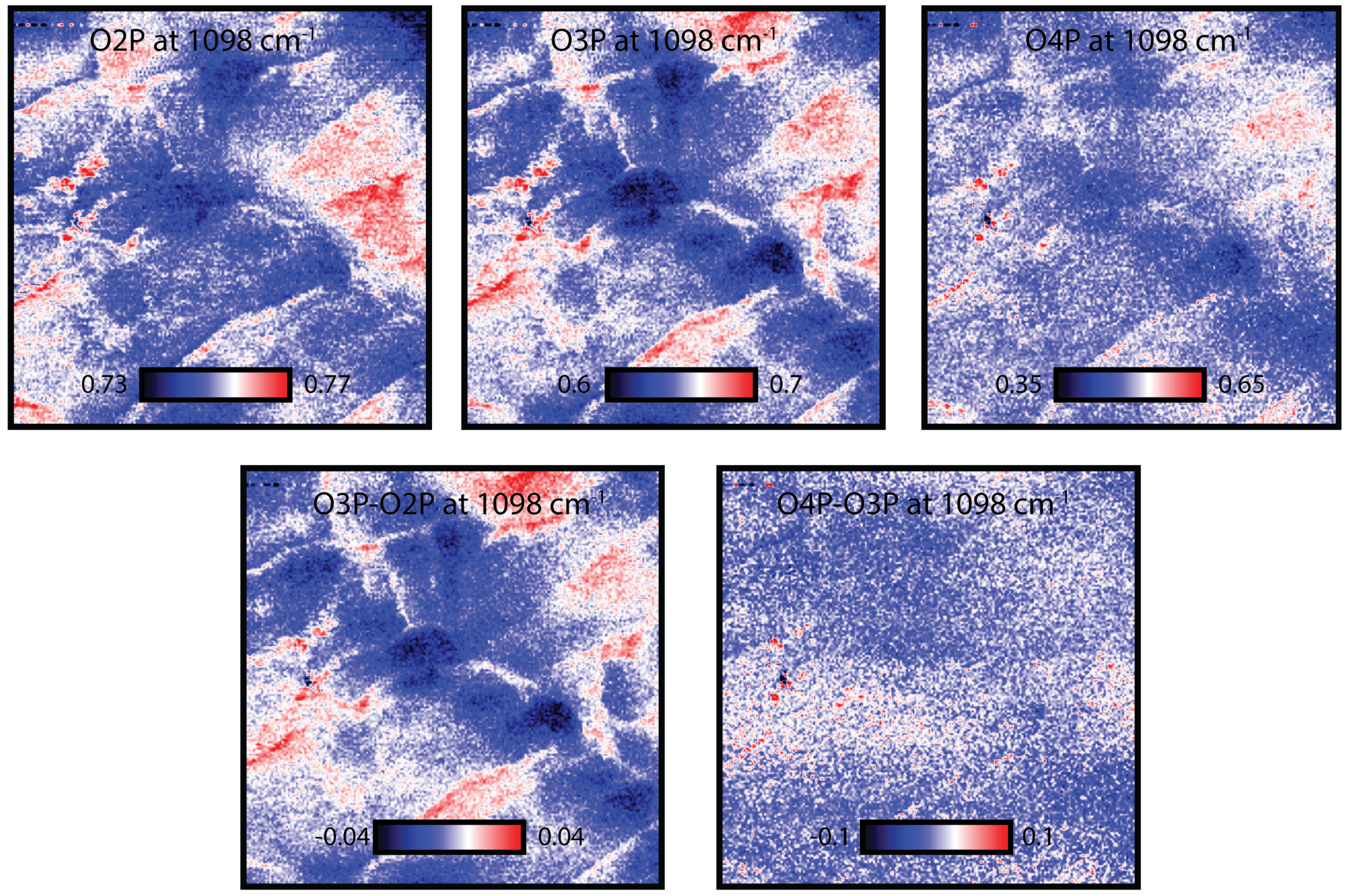

Figure S13. PsHet imaging for $Cu(Im)_2$-310/1 measured under different nearfield optical harmonics.

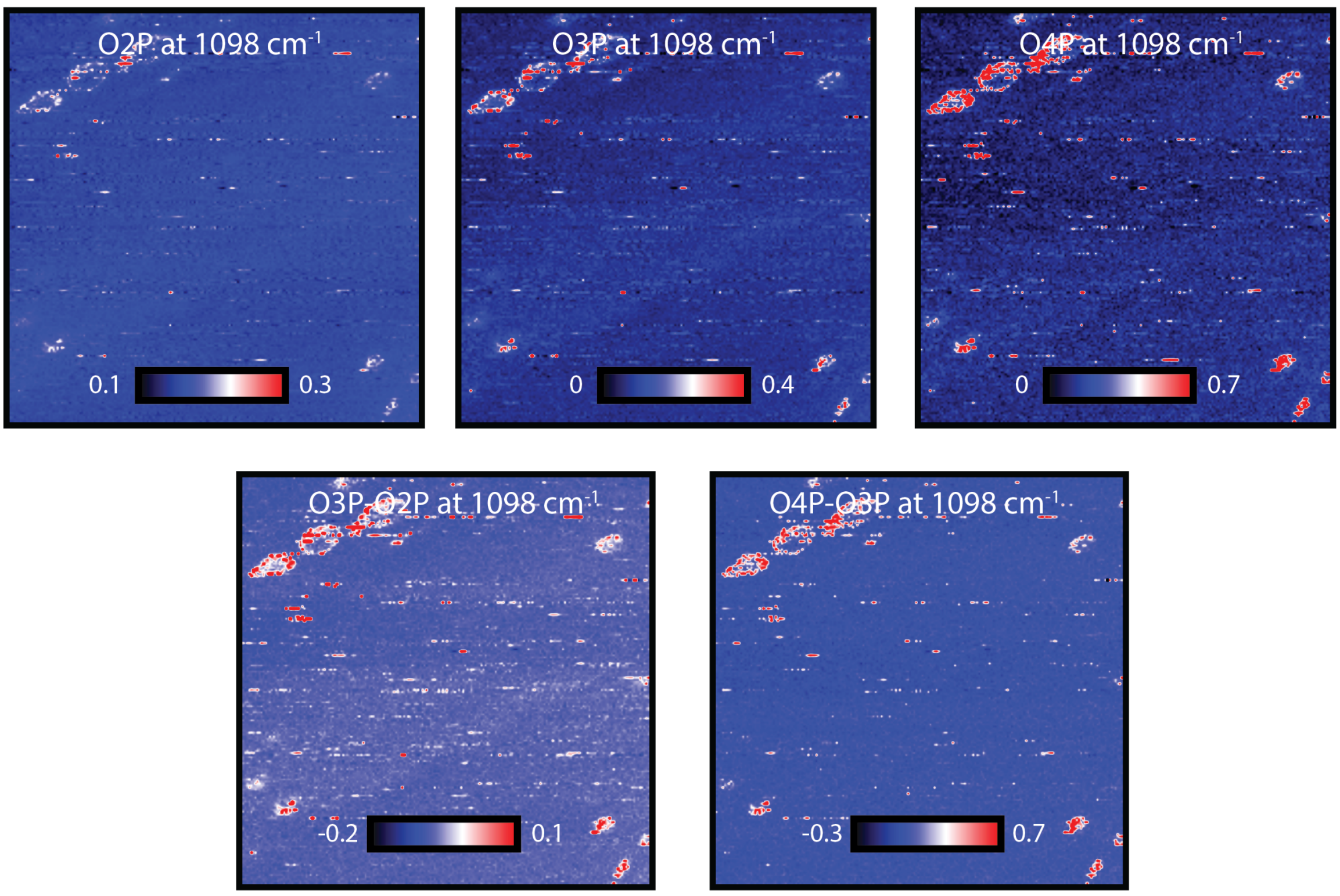

Figure S14. PsHet imaging of $Cu(Im)_2$-245/10 measured under different nearfield optical harmonics.

## 2. Synchrotron methods at Diamond Light Source (B22 MIRIAM)

### 2.1. Far-IR TeraHertz (THz) microspectroscopy (beamtime SM36374)

High-resolution THz infrared microscopy was performed on beamline B22 utilizing the Bruker Vertex 80V FTIR spectrometer in vacuum coupled to a Hyperion 3000 IR microscope, equipped with an *ad hoc* liquid He-cooled Si bolometer. This setup was optimized to measure far-IR THz spectra within the wavelength range below 660 $cm^{-1}$ (as set by a Si beamsplitter in the FTIR and a low-pass filter in the detector). A high-sensitivity liquid-nitrogen cooled metal cadmium telluride (MCT) detector was used in the microscope for the region above 650 $cm^{-1}$ (*via* a KBr beamsplitter and a multilayer filter with a cut-off of above 4000 $cm^{-1}$ in the beam path). Because this study was focused on the THz range microscopy, we employed slits below 500 $\mu$m in size at the sample plane and inserted it into the detected beam path before the bolometer, thereby providing the true spatial resolution for the 1-D line scans and 2-D hyperspectral FTIR scans.

Powder samples were deposited onto a suitable THz-transparent substrate, namely the high resistivity Float Zone Silicon in double-polished disc form (Ø12 mm, thickness 1 mm). The prepared sample holder disks were loaded onto the micro-stage of the Bruker Hyperion 3000 microscope, alongside an identical blank substrate for background measurement. Data collection was performed using the Bruker OPUS v8.5 software.

### 2.2. Far-IR TeraHertz (THz) ATR-FTIR spectroscopy (beamtime SM40142)

THz ATR-FTIR spectroscopy measurements were performed on the Bruker VERTEX 80V FTIR bench with an ATR module, equipped with an *ad hoc* liquid He-cooled Si bolometer. This setup was optimized to measure far-IR THz spectra within the wavelength range below 700 $cm^{-1}$. Pre-synthesized samples of $Cu(Im)_2$ crystals, pellets and glasses were prepared before the beamtime. All samples were pre-evacuated in house to remove residual solvents before beamtime. Measurements were performed under vacuum to below 30 hPa.

The samples were loaded in small quantities (~mg) using a spatula onto the ATR diamond crystal for THz measurements in ATR-FTIR mode. To study the pelleting pressure effect, powdered samples were prepared in a manual hydraulic press fitted with a 13-mm diameter die, where a uniaxial compressive load ranging from 1 to 5 tons was used in preparation of the pellets.

## 3. Volumetric and surface area measurements by nitrogen sorption at 77 K

The analysis was performed on a Quantachrome iQc system. Outgassing of the samples was performed at 120 °C for 12 hours under vacuum. Physisorption measurements were then conducted with liquid $N_2$ at 77 K over the pressure range of $1.4\times10^{-5}$ bar to 1 bar.

Data analysis was performed with Quantachrome supplied software to estimate the surface area. The BET surface areas were computed by fitting the parameters to a straight line over the relative pressure range between 0.05 and 0.2.

Table S1. BET surface areas of different $Cu(Im)_2$ crystals and glasses.

| | $Cu(Im)_2$ | $Cu(Im)_2$-245/10 | $Cu(Im)_2$-310/1 |
|---|---|---|---|
| BET surface areas ($m^2/g$) | 112 | 33 | 19 |

## 4. Further details of nanoindentation methods

The Oliver and Pharr (O&P) method is commonly used to analyze the load-displacement (*P*–*h*) curves obtained from instrumented nanoindentation technique [1, 2]. The O&P method can be used independent of the indenter geometry because it determines the elastic modulus (*E*) based on the assumption of a flat indenter penetrating with a size equal to the contact projected area between the indenter and the sample [3].

Starting from the Sneddon's equation [4], one can write:

$$P = \frac{2E_s h}{1-\nu_s^2}\sqrt{\frac{A}{\pi}}$$

where $P$ is the indenter load, $h$ is the indentation depth, $A$ is the projected area of contact between the flat punch and the sample surface, *E* is the elastic modulus, and ν is the Poisson's ratio. The subscript, s, refers to the sample property.

Denoting $E_r$ as the reduced modulus, i.e. the overall stiffness of the system comprising sample (s) and indenter (i), one gets:

$$\frac{1}{E_r} = \frac{1-\nu_s^2}{E_s} + \frac{1-\nu_i^2}{E_i}$$

The reduced modulus is also given by:

$$E_r = \frac{\sqrt{\pi}}{2}\frac{S}{\sqrt{A}}$$

where $A$ is the projected area of contact between the indenter and the sample surface, $S = \frac{dP}{dh}$ is the contact stiffness between the sample and indenter. The stiffness value *S* may be determined from (a) the slope of the unloading test segment in the *P*–*h* curve beyond the point of maximum load $P_{max}$, or, (b) by the continuous stiffness measurement (CSM) method that continuously oscillates the indenter at a small amplitude (~2-5 nm) and high frequency (~45 Hz) while loading to record these changing values of *S*. The benefit of CSM method is that modulus and hardness values could be determined as a function of indentation depth, instead of only at the single point of unload in the case of (a).

In this study, the indentation modulus $M$ is defined as:

$$M = \frac{E_{\mathrm{s}}}{1 - \nu_{\mathrm{s}}^2}$$

Subsequently, we obtain

$$M = \left[\frac{1}{E_{\mathrm{r}}} - \frac{1 - \nu_{\mathrm{i}}^2}{E_{\mathrm{i}}}\right]^{-1}$$

For a diamond indenter probe, $E_{\mathrm{i}}$ = 1141 GPa and $\nu_{\mathrm{i}}$ = 0.07. When the stiffness of the sample is significantly lower than the indenter stiffness, $E_{\mathrm{s}} \ll E_{\mathrm{r}}$, the second equation can be approximated by:

$$E_{\mathrm{s}} = E_{\mathrm{r}}(1 - \nu_{\mathrm{s}}^2)$$

Because the monolith samples Poisson's ratio is an unknown in this study, we let $\nu_{\mathrm{s}} = 0$, it follows that $E_{\mathrm{s}} \approx E_{\mathrm{r}}$.

The value of the projected area $A$ is found by calibrating an area function in terms of indentation depth $h_c$, $A(h_c)$, by performing indents on a standard material such as fused silica ($E_{\mathrm{s}}$ = 72 GPa and $\nu_{\mathrm{i}}$ = 0.14). The area function is typically expressed as a polynomial of the form:

$$A(h_{\mathrm{c}}) = \sum_{i=0}^{n} C_i h^{(2^{1-i})}$$

where $C_i$ are the fitting coefficients. An area function with three terms were used in this study.

The indentation hardness ($H$), or nanohardness, is defined as

$$H = \frac{P_{\mathrm{max}}}{A|_{h=h_{\mathrm{max}}}}$$

It is important to note that nanohardness is measured when the indenter is under load, which is distinct from other indentation hardness techniques, such as Vickers, where the contact area is determined after unload.

## 5. Fracture toughness, $K_{Ic}$

Table S2. Fracture toughness values for the $Cu(Im)_2$ glasses

| Fracture toughness, $K_{Ic}$ (MPa m$^{½}$) | | $Cu(Im)_2$-310/1 | $Cu(Im)_2$-245/10 |
|---|---|---|---|
| Indenter Hold Time (s) | 1 | 0.491 ± 0.038 | 0.335 ± 0.068 |
| | 10 | 0.442 ± 0.021 | 0.274 ± 0.010 |

Table S3. Fracture toughness values for different MOF materials

| MOF-type materials | Fracture toughness, $K_{Ic}$ (MPa m$^{½}$) | Method | Reference |
|---|---|---|---|
| ZIF-8 monolith (polycrystalline) | 0.074 ± 0.023 | Cube-corner indentation | [5] |
| ZIF-71 monolith (polycrystalline) | 0.16 – 0.22 | Cube-corner indentation | [6] |
| ZIF-71+PMMA monolith | 0.43 – 0.59 | Cube-corner indentation | [6] |
| ZIF-62 glass | 0.104 ± 0.02 | 3-point bending | [7] |
| ZIF-62 glass (Theoretical) | 0.097 ± 0.009 | Molecular dynamics (MD) simulation | [7] |
| $gphen_{0.49}$ZIF-62(Co) glass | 0.103 ± 0.041 | Cube corner indentation | [8] |
| HKUST-1 single crystal | 0.80 ± 0.45 | Micropillar splitting | [9] |

## 6. References

1. Oliver, W.C. and G.M. Pharr, *An Improved Technique for Determining Hardness and Elastic-Modulus Using Load and Displacement Sensing Indentation Experiments*. J. Mater. Res., 1992. **7**: 1564-1583.
2. Oliver, W.C. and G.M. Pharr, *Measurement of hardness and elastic modulus by instrumented indentation: Advances in understanding and refinements to methodology.* J. Mater. Res., 2004. **19**: 3-20.
3. Doerner, M.F. and W.D. Nix, *A method for interpreting the data from depth-sensing indentation instruments*. J. Mater. Res., 1986. **1**: 601-609.
4. Sneddon, I., *The Relation Between Load and Penetration in the Axisymmetric Boussinesq Problem for a Punch of Arbitrary Profile.* Int. J. Engng Sci., 1965. **3**: 47-57.
5. Tricarico, M. and J.C. Tan, *Mechanical properties and nanostructure of monolithic zeolitic imidazolate frameworks: a nanoindentation, nanospectroscopy, and finite element study.* Mater. Today Nano, 2022. **17**: 100166.
6. El Skafi, M., et al., *Mechanical behavior of metal-organic framework monoliths with enhanced fracture toughness.* Mater. Today Nano, 2026. **33**: 100760.
7. To, T., et al., *Fracture toughness of a metal-organic framework glass.* Nat. Commun., 2020. **11**.
8. Weiß, J.-B., et al., *Flux-mediated ligand exchange restructures metal–organic framework glasses.* Nat. Mater., 2026: Article in Press.
9. Tricarico, M. and J.-C. Tan, *Nanostructure-dependent indentation fracture toughness of metal-organic framework monoliths.* Next Materials, 2023. **1**: 100009.